\documentclass[aps,prd,twocolumn,superscriptaddress,showpacs,preprintnumbers,nofootinbib,notitlepage]{revtex4-1}
\usepackage{graphicx}
\usepackage{dcolumn}
\usepackage{bm}
\usepackage{hyperref}
\usepackage{dsfont}
\usepackage{placeins}
\usepackage{todonotes}
\usepackage[utf8]{inputenc}
\usepackage{gensymb}
\usepackage{multirow}
\usepackage{amsmath}
\usepackage{cleveref}
\usepackage{amsfonts}
\usepackage{amssymb}
\usepackage{bbm}
\usepackage{array}   
\newcolumntype{L}{>{$}l<{$}} 
\newcolumntype{R}{>{$}r<{$}}
\newcolumntype{C}{>{$}c<{$}}
\usepackage{color}
\usepackage{xcolor}
\usepackage{xspace}
\usepackage{braket}
\usepackage{units}
\usepackage{slashed}
\usepackage{multirow}
\usepackage[normalem]{ulem}
\hypersetup{ 
    pdfnewwindow=true,      
    colorlinks=true,       
    linkcolor=blue,          
    citecolor=blue,        
    filecolor=blue,      
    urlcolor=blue        
}

\graphicspath{ {plots/} }

\newcommand{\be}{\begin{equation}}
\newcommand{\ee}{\end{equation}}

\newcommand{\im}{\mathrm{Im}\,}
\newcommand{\re}{\mathrm{Re}\,}
\newcommand{\e}{\mathrm{e}}

\newcommand{\IC}{{\Bbb C}}

\usepackage{bm} 

\newcommand{\mevnospace}{\ensuremath{{\mathrm{\,Me\kern -0.1em V}}}}
\newcommand{\gevnospace}{\ensuremath{{\mathrm{\,Ge\kern -0.1em V}}}}
\newcommand{\tevnospace}{\ensuremath{{\mathrm{\,Te\kern -0.1em V}}}}
\newcommand{\mev}{\mevnospace\xspace}
\newcommand{\gev}{\gevnospace\xspace}

\newcommand{\sig}
{\ensuremath{\sigma/f_0(500)}\xspace}

\newcommand{\fzero}
{\ensuremath{f_0(980)}\xspace}

\newcommand{\pipi}
{\ensuremath{\pi \pi \to \pi \pi}\xspace}

\usepackage{mfirstuc} 
\newcommand{\addReviewer}[2]{
  \expandafter\newcommand\csname #1\endcsname[1]{{\bf \color{#2} \capitalisewords{#1}:\,##1}}
  \expandafter\newcommand\csname #1cor\endcsname[2]{{\color{#2} \capitalisewords{#1}:\,\st{##1}{\,\bf ##2}}}
  \expandafter\newcommand\csname #1color\endcsname{\,#2}
}

\usepackage{tikz, marginnote} 
\newcommand{\checkedby}[1]{
\ifdefined\CROSSCHECKS
  \marginnote{
    \begin{tikzpicture}
      \foreach \x [count=\xi] in {#1} {
         \node[shape=circle,inner sep=0mm,
         minimum size=2mm,
         fill=\csname \x color\endcsname] at (\xi*3mm,0) {};
       }
    \end{tikzpicture}
  }
\else
\fi
}

\usepackage{soul,color}
\definecolor{chromeyellow}{rgb}{1.0, 0.65, 0.0}
\definecolor{DodgeBlue}{rgb}{0.118, 0.565,1.000}
\definecolor{asparagus}{rgb}{0.53, 0.66, 0.42}
\definecolor{cadmiumgreen}{rgb}{0.0, 0.42, 0.24}

\addReviewer{AR}{asparagus}
\addReviewer{JR}{red}
\addReviewer{JRE}{blue}

\newcommand{\ucm}{Departamento de F\'isica Te\'orica and IPARCOS, 
Universidad Complutense de Madrid, 
E-28040 Madrid, Spain}
\newcommand{\bern}{Albert Einstein Center for Fundamental Physics, Institute for Theoretical Physics,
University of Bern, Sidlerstrasse 5, 3012 Bern, Switzerland}

\begin{document}
\title{Global parameterization of $\pi \pi$ scattering up to 2 \gev}
\author{J.~R.~Pelaez}
\email{jrpelaez@fis.ucm.es}\affiliation{\ucm}
\author{A.~Rodas}
\email{arodas@ucm.es}\affiliation{\ucm}
\author{J.~Ruiz de Elvira}
\email{elvira@itp.unibe.ch}\affiliation{\bern}
\begin{abstract}
We provide global parameterizations of \pipi scattering $S0$ and $P$ partial waves up to 
roughly 2 GeV for phenomenological use. These parameterizations describe the output
and uncertainties of previous partial-wave dispersive analyses of \pipi, both in the real axis up to 1.12 \gev and in the complex plane within their applicability region, while also fulfilling forward dispersion relations up to 1.43 \gev. Above that energy we just describe the available experimental data. Moreover, the analytic continuations of these global parameterizations also describe accurately the dispersive determinations of the
\sig, \fzero and $\rho(770)$ pole parameters.
\end{abstract}

\maketitle

\section{Introduction} 

The unprecedented high statistics on hadronic observables attained at experiments like LHCb, Belle or BaBar require rigorous and precise parameterizations of final state interactions. Future Hadronic facilities (FAIR, PANDA, etc..) will be even more demanding. 
One of the most needed parameterizations is that of \pipi scattering, since two or more pions appear very frequently as final products of many hadronic interactions. In addition, a renewed interest on \pipi scattering is coming from lattice calculations, which have been recently able to obtain scattering partial waves
with almost realistic masses \cite{Briceno:2017max}.

Data on \pipi scattering were obtained in the 70's~\cite{Hyams:1973zf,Grayer:1974cr,Hyams:1975mc,Kaminski:1996da,Protopopescu:1973sh} indirectly from the $\pi N\rightarrow \pi\pi N'$ reaction. Unfortunately, this technique gave rise to several conflicting data sets. Thus, for decades, crude models were enough to describe such data.
The exception is the very low-energy region, both in the experimental and theoretical fronts. On the one hand, there are very precise data below the kaon mass coming from $K_{l4}$ decays~\cite{Rosselet:1976pu,Pislak:2001bf}, particularly after the NA48/2 results~\cite{Batley:2010zza}.  On the other hand, Chiral Perturbation Theory (ChPT) \cite{Gasser:1983yg,Gasser:1984gg} provides a systematic and accurate low-energy expansion in terms of pion masses and momenta. 

However, for most phenomenological applications the low-energy region is not enough, since the production of pions is generically more copious around resonances.
ChPT can be successfully extended to the resonance region by means of dispersion relations \cite{Truong:1988zp,Dobado:1989qm,Dobado:1992ha,Dobado:1996ps}, usually called Unitarized ChPT. Different versions or approximations of this method generate or reconstruct all resonances in \pipi up to 1.2 GeV: the $\sigma/f_0(500)$ the $\rho(770)$ and the $f_0(980)$ \cite{Oller:1997ti,Oller:1998hw,GomezNicola:2001as,Pelaez:2004xp,Nieves:1999bx} and even in $\pi K$ scattering. However, the prize to pay is the loss of a controlled systematic expansion, which hinders the calculation of uncertainties, and the length of the analytic expressions once one deals with coupled channels above $K \bar K$ threshold. Above 1.2 GeV one can introduce by hand other resonances, yielding a successful description of data \cite{Oller:1998zr}, although with the same caveats as before and with expressions even more elaborated.
Nevertheless, the interest of these unitarized approaches is that they can connect with QCD through the chiral parameters. Moreover, they provide a good semi-quantitative approximation, including values of resonance poles, which are much better than the usual description of two-pion interactions in terms of simple popular models, like the superposition of simple resonant shapes, Breit-Wigner formulas in different versions, isobar models, etc...For a recent review on modern $\pi\pi$ scattering dispersive determinations, UChPT and other models for the \sig see~\cite{Pelaez:2015qba}.

The interest of those naive popular models is, on the one hand, their simplicity, since for most applications just the phase and the elasticity functions are needed, not an elaborated model of the interactions with other channels. On the other hand, they can be fairly reasonable for narrow isolated resonances, like the $\rho(770)$. However, such simple models provide an incorrect description of the scalar-isoscalar partial wave. In particular this is the case of the very broad $\sigma/f_0(500)$ pole and its interplay with the very narrow $f_0(980)$, together with the singularity structure in terms of cuts in the complex $s$ plane. Actually, the rescattering of two pions in this channel is frequently described with some sort of Breit-Wigner parameterization for the $\sigma/f_0(500)$, 
which might be able to describe a wide bump in the data, but fails to describe  the chiral constraints in the threshold region as well as the phase shift in the whole $\sigma/f_0(500)$ region. Recall that by Watson's Theorem~\cite{Watson:1954uc} any strong elastic rescattering of two pions must have the very same phase of the \pipi partial-wave with the same isospin and angular momentum.

In general, modern Hadron Physics demands more precise and model-independent meson-meson scattering parameterizations. This has been achieved over the last two decades by means of dispersion relations aiming at precision, not only for $\pi\pi$~\cite{Ananthanarayan:2000ht,Colangelo:2001df,DescotesGenon:2001tn,Kaminski:2002pe,GarciaMartin:2011cn,Kaminski:2011vj,Moussallam:2011zg,Caprini:2011ky,Albaladejo:2018gif}, but also for $\pi N$~\cite{Ditsche:2012fv,Hoferichter:2015hva} or $\pi K$ scattering~\cite{Buettiker:2003pp}. Unfortunately, we have found that, for the hadron community, these dispersive results, either obtained numerically from complicated integral equations or parameterized by piecewise functions, are not always so easy to implement or do not cover a sufficiently large energy region. 
Hence, the purpose of this work is to provide relatively simple and ready-to-use parameterizations of the phase and elasticity of the scalar-isoscalar and vector \pipi scattering partial waves up to almost 2 GeV.  They will be consistent with data 
globally from threshold up to approximately 2 GeV, and with the dispersive analysis in~\cite{GarciaMartin:2011cn}, which extends up to 1.43 GeV in the real axis. Moreover, we will impose that these parameterizations
will provide a simple analytic continuation to the complex plane, consistent with the dispersive representation and the values for the pole positions and residues of the $\sigma/f_0(500)$, $\rho(770)$ and $f_0(980)$ resonances found in~\cite{GarciaMartin:2011jx}. In addition, both the dispersive results for the threshold and subthreshold regions are also described by the global parameterization, consistently with the scattering lengths, slope parameters and $S0$ wave Adler zero values obtained in~\cite{GarciaMartin:2011cn}.

\section{The input to be described}
\label{sec:input}

As we already commented, there are several \pipi scattering data sets extending up to almost 2 GeV~\cite{Hyams:1973zf,Protopopescu:1973sh,Grayer:1974cr,Hyams:1975mc,Kaminski:1996da}. 
These are customarily given in terms of partial waves $t^I_\ell$ of definite isospin $I$ and angular momentum $\ell$. We will also use the spectroscopic notation where the $\ell=0,1,2,3...$ waves are referred to as S, P, D, F... waves, followed by their isospin.
Unfortunately, all those data sets are often incompatible from one another and, moreover, simple fits to each separated set or to averaged data sets do not satisfy well dispersion relations~\cite{Pelaez:2004vs,Kaminski:2006qe,Kaminski:2006yv,Yndurain:2007qm,Kaminski:2006qe, GarciaMartin:2011cn}. Nevertheless, it is possible to use dispersion relations as constraints to obtain a Constrained Fit to Data (CFD)~\cite{GarciaMartin:2011cn} that still describes the \pipi data on partial-waves but satisfies dispersion relations within uncertainties. Furthermore, the CFD fulfills the normality requirements of the residual distribution~\cite{Perez:2015pea}, hence ensuring that the standard
approach for error propagation can be used. This CFD parameterization will thus be part of our input.

One might wonder why not using directly this CFD parameterization and why in this work we are trying to obtain another one. After all, this parameterization has become quite popular and it has been used in many phenomenological applications. There are several reasons.

First, the dispersion relations used in~\cite{GarciaMartin:2011cn} are of two kinds and they were applied up to different energies, always below 2 GeV. One kind consists of a set of Forward Dispersion Relations, which were studied up to 1.43 GeV. These equations are rather simple, but unfortunately cannot be extended to the complex plane in search for poles. They are only useful as constraints on the real axis. The other kind consists of two sets of partial-wave dispersion relations, usually referred to as Roy equations~\cite{Roy:1971tc,Basdevant:1972uu,Basdevant:1972uv,Basdevant:1973ru,Pennington:1973xv,Pennington:1973hs,Pennington:1974kp,Ananthanarayan:2000ht,Colangelo:2001df,Moussallam:2011zg,Caprini:2011ky} (with two subtractions) and GKPY equations~\cite{GarciaMartin:2011cn} (with one subtraction). The former are more stringent in the low-energy region and the latter in the resonance region. Unfortunately, these partial-wave  equations are limited to 1.12 GeV, although they can be rigorously continued to the complex plane in search for resonance poles. The existence of these different energy regions motivated the authors in~\cite{GarciaMartin:2011cn} to describe the data with a piecewise parameterization, which in principle cannot be extended rigorously to the complex plane.
Therefore, our first aim is to provide a rather simple but global analytic parameterization, with realistic uncertainties, that can be used from $s=0$ to 1.43 GeV. Thus, it will mimic the CFD piecewise parameterization {\it in the real axis} above the $\pi\pi$ threshold, which will be used as the first of our inputs to be described.

Second, the $\sigma/f_0(500)$ pole lies so deep in the complex plane that a careful dispersive determination is needed in order to extract its precise parameters rigorously~\cite{Caprini:2005zr,GarciaMartin:2011jx,Moussallam:2011zg}. Using the CFD parameterization as input in the GKPY equations, it was obtained numerically that its pole lies at $\sqrt{s_\sigma}=(457^{+14}_{-13})-i(279^{+11}_{-7})\mev$ with a residue $\vert g\vert=3.57^{+0.11}_{-0.13}$. Now, the low-energy piece of the CFD parameterization~\cite{GarciaMartin:2011cn} was constructed as a conformal expansion valid up to 850 MeV, which lies within the elastic \pipi region. This CFD conformal piece can be continued to the complex plane finding $\sqrt{s_\sigma}=(474\pm 6)-i(254\pm4)\mev$, which is fairly close, but it is not the pole obtained from the dispersive representation. 
This discrepancy does not improve when one includes further constraints in the real axis. Namely, even if the CFD conformal parameterization is extended up to the $K\bar K$ threshold to take into account the $f_0(980)$ effect or to the subthreshold region, in order to describe the dispersive value for the Adler zero, one still finds sizable discrepancies with the GKPY pole result. This problem was observed time ago~\cite{Caprini:2008fc,Masjuan:2014psa,Caprini:2016uxy,Pelaez:2016klv}; arbitrarily small changes in the real axis input data may lead to indefinitely large variations for the analytic continuation to the complex plane. This illustrates how trying to obtain the $\sig$ pole from a data fit 
that only reaches 850 MeV is not precise enough. Actually, the effects of the $f_0(980)$ and other singularities, like the left hand cut, are significant at this level of precision. Hence, our second aim is to provide a simple analytic parameterization that 
reproduces simultaneously the dispersive poles of the $\sig$ and $f_0(980)$ and their interference. Thus, the numerical results of the GKPY dispersion relations in the complex plane, including the numerical values of the $\sig$ and $f_0(980)$ poles, and in the subthreshold region will be the second input to be described. For the $P$-wave we will proceed similarly, but just for the $\rho(770)$ pole.

Finally, the CFD parameterization and the dispersive data analysis from which it was obtained only reach 1.43 GeV, but there are more data up to almost 2 GeV. However, the data at those high energies have many well-known caveats. Some of them were already discussed in detail in~\cite{Pelaez:2003eh,Pelaez:2015qba} and in appendix C of~\cite{Pelaez:2004vs}, but we summarize them here.
First, in that energy region we have to rely on a single scattering experiment, the CERN-Munich Collaboration, so that systematic uncertainties relative to other experiments are not available. Nevertheless there is some information on partial waves for the final $\pi^0\pi^0$
system above 0.8 GeV from the GAMS experiment on $\pi^-p\rightarrow\pi^0\pi^0 n$ \cite{Alde:1998mc}, which will be of interest later. Second, the CERN-Munich collaboration has many different solutions for the $\pi\pi$ scattering partial waves.
Of these, the most popular one for the S0-wave is the one published in 1973~\cite{Hyams:1973zf}, also called ``solution b'' in the collaboration compilation of Grayer et al.~\cite{Grayer:1974cr}. This solution is also consistent with a later reanalysis with polarized targets~\cite{Kaminski:1996da}. In addition, there is the ``solution (- - -)'', which was the most favored in the 1975 collaboration reanalysis~\cite{Hyams:1975mc} and the most used solution for the P-wave.
Note that both ``b'' and ``(- - -)'' solutions are compatible with one another below 1.43 GeV.
Other solutions for both waves were already disfavored in that very same analysis. 
Third, both solutions have caveats. On the one hand, the inelastic contribution to all hadronic cross sections are expected to dominate over the elastic ones (something that has been verified for $\pi N$, $KN$ and $NN$ scattering). However, this is not the case of ``solution b''. It is hard to understand why this should be different for pions. On the other hand, if the inelasticity is large, then it can be proved theoretically~\cite{Atkinson:1969wy,Atkinson:1970pe} that the solution in terms of phase and elasticity is not unique. ``Solution b'' is an example of an almost elastic case and ``solution (- - -)'' of a strong inelastic effect. Finally, the very same convergence of the partial-wave expansion could be questioned at those energies, since around 1.7 GeV the F-wave is as large as the P-wave, the D0 wave as large as the S0 and the D2 actually larger than the S2.
Furthermore, the solution (- + -) has been recently revisited and slightly modified by one of the authors of the CERN-Munich Collaboration \cite{Ochs:2013gi}. He argues that it is still consistent if the I=2 wave is not considered elastic, finding some qualitative agreement with the 
GAMS experiment. In particular they both show hints of the $f_0(1500)$ resonance,
in contrast to the other solutions. Note that this (- + -) solution is also compatible with the "b" and (- - -) solutions below 1.43 GeV.

Therefore, in view of the caveats above, we have extended our fits beyond 1.43 GeV using as our third source of input different sets of data. Namely: I) the data of~\cite{Hyams:1973zf,Grayer:1974cr,Kaminski:1996da}, which reaches up to 1.9 \gev,
to obtain a ``solution I'', or II) the (- - -) data of~\cite{Hyams:1975mc}, which reaches up to 1.8 \gev, to obtain a ``solution II'', or III) the updated solution (- + -) in \cite{Ochs:2013gi} to obtain a ``solution III".
Below 1.43 GeV the input is the same for all solutions and they agree within uncertainties.
As a technical remark, we have ensured that the central value and the first derivative of both the phase and elasticity are continuous at the matching point,
which is chosen at 1.4 GeV to avoid fitting the very end of the CFD parameterization.
In any case, one should keep in mind that none of these solutions has been checked against dispersion relations above 1.43 GeV. Thus, beyond that energy they should be considered purely phenomenological data fits.

\section{Analytic parameterizations}\label{Sec:parameterizations}

In this section we present the parameterizations used for the scalar-isoscalar and vector \pipi partial waves. 
Let us first note that below the $K \bar K$ threshold the process will be considered elastic and hence it will be uniquely characterized by its phase shift $\delta^I_\ell(s)$, as it is customary, through the following definition:
\begin{equation}\label{eq:elastict}
t^I_\ell(s)=\frac{\hat t^I_ \ell(s)}{\sigma(s)}=\frac{e^{i\delta^I_\ell(s)}\sin\delta^I_\ell(s)}{\sigma(s)}=
\frac{1}{\sigma(s)}\frac{1}{\cot\delta^I_\ell(s)-i},
\end{equation}
where $\sigma_i(s)= 2 q_i(s)/\sqrt{s}=\sqrt{1-4m_i^2/s}$ is the two-body phase space, $q_i(s)$ being the CM momentum of a particle with mass $m_i$. For brevity we will suppress the subindex in the pion case and write $\sigma(s)\equiv\sigma_\pi(s)$. 
The elastic region will be described with conformal maps for both the $S$ and $P$ waves.

Let us also recall the standard inelastic partial-wave representation
\begin{equation}
    t^I_\ell(s)=\frac{\eta^I_\ell(s) \e^{2i \delta^I_\ell(s)}-1}{2i \sigma(s)},
\end{equation}
where the elasticity parameter $\eta^I_\ell(s)$ and phase shift $\delta^I_\ell(s)$ will be described by two independent real functions. 

Let us now present separately the parameterizations we have used 
to describe the two partial waves of interest for this work.

\subsection{S0-wave parameterization}

As explained above, our parameterizations will be consistent with the dispersive data analysis of~\cite{GarciaMartin:2011cn}, which extends up to 1.43 \gev. Above that we will only provide three phenomenological fits to three sets of incompatible data, carefully matched to our parameterizations below. Let us discuss both regions separately.

\subsubsection{S0-wave parameterization below 1.4 \gev}

The \sig and \fzero resonances dominate the behavior of the S0 partial wave in this region. The somewhat controversial $f_0(1370)$ couples very weakly to two pions and its effect in this region can be treated as background. For our purposes it is important to remark that the \sig has an associated pole very  deep in the complex plane that produces a wide structure increasing monotonously from threshold up to roughly 900 MeV, reaching a phase-shift of $90\degree$ around 800 MeV, as seen in Fig.~\ref{fig:shift-inelas}. 
It is known~\cite{Yndurain:2007qm} that the \sig pole can be generated in the S0 partial wave by a simple truncated conformal expansion, that we will call $t^0_{0,\text{conf}}(s)$. However, above 900 MeV the \fzero pole adds a further sharp increase that makes the phase larger than $200\degree$ right below the $K\bar K $ threshold. The phase then keeps growing slower but monotonously up to 2 GeV.

It is worth noticing that the interplay between the \sig and \fzero poles produces a sharp dip in the modulus of the amplitude and the elasticity right above the $K\bar K $ threshold.
 In order to describe the $f_0(980)$ effects accurately and consistently with the dispersive results, we will factorize in the $S$ matrix the \fzero shape separately from the conformal expansion that contains the \sig pole. In other words, we will use $S^0_0=S^0_{f_0}S^0_{0,\text{conf}}$, where
 \begin{eqnarray}
      S^0_0&=&1+2i\sigma(s)t^0_0\,,\nonumber\\ S^0_{0,\text{conf}}&=&1+2i\sigma(s)t^0_{0,\text{conf}}\,, \nonumber\\ S^0_{f_0}&=&1+2i\sigma(s)t^0_{f_0}\,.
 \end{eqnarray}
This factorization ensures elastic unitarity for the S0 wave, i.e., $\vert S^0_0\vert=1$, when both the conformal and the $f_0(980)$ contributions fulfill elastic unitarity independently, i.e.,  $\vert S^0_{0,\text{conf}}\vert=\vert S^0_{f0}\vert=1$. This will be the case in the elastic region below the $\bar KK$ threshold, $s<4m_K^2$.

For our purposes, we are interested in the amplitude partial-wave
\begin{equation}
t^0_0(s)=t^0_{0,\text{conf}}(s)+t^0_{f_0}(s)+2i\sigma(s)t^0_{0,\text{conf}}(s)\,t^0_{f_0}(s).
\label{eq:finalpa}
\end{equation}

Now, the conformal factor of the partial wave is constructed by analogy to the elastic formulation in Eq.~\eqref{eq:elastict}
\begin{equation}\label{eq:tconf}
t^0_{0,\text{conf}}(s)=
\frac{1}{\sigma(s)}\frac{1}{\Phi^0_0(s)-i}, \quad s<1.4 \gev
\end{equation}
where, building on~\cite{Yndurain:2007qm}
\begin{align}
\Phi^0_0(s)=\frac{\sqrt{s}}{2q(s)}\frac{m_\pi^2}{s-z_0^2/2}\left(\frac{z_0^2}{m_\pi \sqrt{s}}+\sum_{n=0}^N{B_n \omega(s)^n}\right).
\label{eq:generalconformalS0}
\end{align}
Let us remark that $\Phi^0_0(s)=\cot\delta^0_0(s)$ in the elastic region $s\leq 4m_K^2$,
implying $\vert S^0_{0,\text{conf}}\vert=1$. The $s-z_0^2/2$ denominator provides the so-called Adler zero at $s_{Adler}=z_0^2/2$ required by chiral symmetry~\cite{Adler:1964um}. For $z_0=m_\pi$ one would recover the Current-Algebra result, namely the leading order ChPT value. However, for us it will be a free parameter, fundamental to describe the subthreshold region. As we will see, it comes out from the fits consistent with the dispersive evaluation, which in turn is consistent with higher order ChPT evaluations (see~\cite{GarciaMartin:2011cn,Nebreda:2012ve}).
Note also that if we did not include the $\sim1/\sqrt{s}$ term added to the conformal series then, for the values of $B_n$ obtained from the fit, we would find in the unphysical region 
that $\Phi^0_0(s_{Adler})\rightarrow +i\infty$ and $\Phi^0_0(0)\rightarrow i0^+$,
so that necessarily $\Phi^0_0(s)=i$ somewhere between those two subthreshold points. This would yield a spurious pole, i.e., a ghost, in the partial wave.
As explained in~\cite{Yndurain:2007qm} these ghosts are mostly harmless and have little relevance in the fit quality and the pole positions, but as a matter of principle it is better to remove them.
Actually, as explained there, adding the $1/\sqrt{s}$ term to the conformal series simply shifts the value $\Phi^0_0(0)$ so that the spurious pole disappears. The $1/\sqrt{s}$ cut does not introduce any additional singularity in the partial wave since its square-root cut falls right on the left cut. At the same time this term is suppressed in the physical region and thus barely affects the fit and resonance pole positions.   
As shown below, for this wave it will be enough to set $N=5$ to obtain a good overall $\chi^2/d.o.f.$ in the elastic region.

The conformal variable is defined as
\begin{equation}
    \omega(s)=\frac{\sqrt{s}-\alpha \sqrt{s_0-s}}{\sqrt{s}+\alpha \sqrt{s_0-s}},
    \label{eq:coformalvariable}
\end{equation}
where $s_0$ corresponds to the highest value of $s$ where the expansion is real and then $\alpha$ sets the center of the conformal expansion. 
We have found in practice that the S0 wave is more conveniently described if the conformal expansion, by becoming imaginary, introduces some inelasticity above the $K \bar K$ threshold~\cite{Caprini:2008fc}). 
Thus, we choose $s_0= 4 m_K^2$ with $\alpha=1$ for simplicity, so that the expansion center lies near 0.7 \gev. Hence, between $K\bar K$ threshold and 1.4 \gev,
the $\Phi^0_0(s)$ function will be complex, which effectively introduces an inelasticity.
We tried otherwise, with a higher $s_0$, so that no inelasticity would come from the conformal factor.
We then found that the $t^0_{f_0}(s)$ would require many extra parameters with strong correlations. In addition these parameters would have huge and unnatural scale differences among themselves.

Actually, with our choice of conformal expansion we 
can use a simple and intuitive functional form for $t^0_{f_0}$, inspired by the expression used in~\cite{Caprini:2008fc}. Namely
\begin{align}
    t^0_{f_0}(s)=\frac{s \,G }{M-s-\bar{J} (s,m_\pi)\,s\, G -\bar{J}(s,m_K) m_K^2 f(s)},
\label{eq:ine}
\end{align}
which ensures elastic unitarity for $s<4m_K^2$. 
The $\bar J$ two-meson loop functions, or Chew-Mandelstam functions \cite{Chew:1960iv}, provide the unitarity 
cut above each threshold and are defined as
\begin{equation}
\bar{J}(s,m_i)=\frac{2}{\pi}+\frac{\sigma_i(s)}{\pi}\log\left(\frac{\sigma_i(s)-1}{\sigma_i(s)+1}\right).
\label{eq:Jbar}
\end{equation}
Note that the constant $G$ in the numerator in Eq.~\eqref{eq:ine} is multiplied by $s$ in order to cancel the phase-space pole at $s=0$ in Eq.~\eqref{eq:finalpa}. In addition, this factor suppresses the inelastic contribution at low energies, hence ensuring that the low-energy region is dominated by the conformal parameterization. In principle, $f(s)$ could be any real analytic function and for convenience we will build it  as an expansion of Chebyshev polynomials. The main advantage of this procedure is the low correlation among parameters, which will provide a more realistic description of the uncertainties. Note that the expansion variable will not be $s$, but a linear transformation that maps the  $[2 m_K,1.5\gev]$  energy region into the $[-1,1]$ segment, where Chebyshev polynomials are orthogonal. 
This variable is:
\begin{equation}
    \omega_1(s)=2\frac{\sqrt{s}-2m_K}{1.5\gev-2m_K}-1.
\end{equation}
Thus, the real function $f(s)$ will be expanded as
\begin{equation}
    f(s)=\sum_{i=0}^N K_i x_i(\omega_1(s)),
     \label{eq:chebexp}
\end{equation}
where $x_i$ is the Chebyshev polynomial of order $i$ and $K_i$ are fitting constants. In practice it is enough to set $N=3$ and so we will do. This function also suppresses the \fzero contribution far from its nominal mass.

As a matter of fact, one can get an acceptable $\chi^2/d.o.f.$ using Eq.~\eqref{eq:ine} to fit the dispersive results in the real axis and the complex plane around the $K\bar K$ threshold. However, when so doing the \fzero pole position does not come out at the precise dispersive value given in~\cite{GarciaMartin:2011cn}. For this reason
we will impose the dispersive value of its pole position in the fit, by fixing the $G$ and $M$ constants.

Let us then briefly recall how to reach the second Riemann sheet in search for poles. According to the $S$-matrix unitary relation $S S^{\dagger}=\mathbbm{1}$ and taking into account the Schwartz reflection symmetry,  $t^I_\ell(s+i \epsilon)=t^I_\ell(s-i \epsilon)^*$, then the partial wave in the second Riemann sheet $t_{l}^{(2),I}$ is algebraically related to itself in the first Riemann sheet by
\begin{equation}
    t^{(2), I}_ \ell(s)=\frac{t^{I}_\ell(s)}{1+2i \sigma^{}(s) t^{I}_\ell(s)},
    \label{eq:2rs}
\end{equation}
where the $\sigma(s)$ determination is chosen so that $\sigma(s^*)=-\sigma(s)^*$ to ensure the Schwartz reflection symmetry
\begin{equation}
    t^{(2), I}_ \ell(s^*)=t^{(2), I}_ \ell(s)^*.
\end{equation}
As a consequence, a pole $s_p=s_R+i s_I$ in the second Riemann sheet implies that a zero of the $S$ matrix exists also in the first Riemann sheet at $s_p$. This imposes two constraints on Eq.~\eqref{eq:ine},
which allow us to fix $G$ and $M$ as follows:

\begin{widetext}
\begin{align}
    M&=\frac{(f_I J_{K R}+f_R J_{K I}) (s_I (J_{\pi I}-2 \sigma_R)-s_R (J_{\pi R}+2 \sigma_I))+(J_{\pi I}-2 \sigma_R)
   \left(s_I^2+s_R^2\right)}{d} -(f_I J_{K I}-f_R J_{K R}), \nonumber \\
    G&=-\frac{f_I J_{K R}+f_R J_{K I}+s_I}{d},
\label{eq:MGcons}
\end{align}
\end{widetext}
where we have defined the constants
\begin{align}
 f(s_p)&=f_R+i\,f_I, \nonumber \\
 \bar{J}(s_p,m_K)&=J_{K R}+i\,J_{K I}, \nonumber \\
 \bar{J}(s_p,m_\pi)&=J_{\pi R}+i\,J_{\pi I}, \nonumber \\
 \sigma(s_p)&=\sigma_R+i\,\sigma_I, \nonumber \\
d&=J_{\pi I} s_R+J_{\pi R} s_I+2 (\sigma_I s_I- s_R \sigma_R), 
\end{align}
and  $s_{p}=s_R+i s_I$ corresponds to the \fzero pole position, which is therefore a parameter to be varied within its uncertainties in our formulas.

In summary, for the scalar wave  below $1.4$\gev we will use Eq.~\eqref{eq:finalpa} with $t^0_{f_0}(s)$ defined in Eqs.~\eqref{eq:ine}, \eqref{eq:Jbar} and ~\eqref{eq:MGcons},
whereas $t^0_{0,\text{conf}}(s)$, containing the \sig pole,
is defined in Eqs.~\eqref{eq:tconf} and \eqref{eq:generalconformalS0},
which above the $K \bar K$ threshold gives and additional contribution to the inelasticity besides that of $t^0_{f_0}$.

\subsubsection{S0-wave parameterization above 1.4 \gev}

As we have emphasized repeatedly, from 1.43 \gev there are no dispersive data analyses and, besides, the data can be grouped into three inconsistent data sets. However, we are frequently asked if we could extend our parameterization beyond 1.43 \gev. Thus, we will provide simple phenomenological fits to the three sets of data that we will match to our formulas below 1.4 GeV so that the whole parameterization and its derivative are continuous. 
For this we need the values at $s_m=(1.4 \gev)^2$ of the phase shift, the elasticity and their derivatives with respect to the energy squared, denoted with a prime. These inputs will be taken from the parameterizations below 1.4 \gev.

To reduce the number of parameters, we will make use again of Chebyshev polynomials to describe the phase shift above $s_m$, 
namely
\begin{align}
    \delta^0_0(s)&=\delta^0_0(s_m)+\Delta^0_0 [x_1(\omega_2(s))+1]+d_0 [x_2(\omega_2(s))-1]. \nonumber \\ 
    &+d_1 [x_3(\omega_2(s))+1]+d_2 [x_4(\omega_2(s))-1],\nonumber \\
    \Delta^0_0&=\frac{\delta^{\prime\,0}_0(s_m)}{\omega’_2(s_m)}-d_0 x'_2(-1) -d_1 x'_3(-1)-d_2 x'_4(-1), \nonumber\\
    &=\frac{\delta^{\prime\,0}_0(s_m)}{\omega’_2(s_m)}+4d_0-9d_1+16d_2.\label{eq:highphase00} 
\end{align}
 The variable for the Chebyshev polynomials now is:
\begin{align}
    \omega_2(s)&=2\frac{\sqrt{s}-\sqrt{s_m}}{2 \gev-\sqrt{s_m}}-1. \nonumber 
\end{align}
The presence of $\delta^0_0(s_m)$ in Eq.~\eqref{eq:highphase00} ensures the continuity of the parameterization 
and the value of $\Delta^0_0$ the continuity of the derivative.
As we will see when fitting the data, we will need just two Chebyshev polynomials to get a good $\chi^2/d.o.f.$ for solutions I and II, leaving just 
one free parameter, $d_0$, for those fits. However, three free parameters are needed to obtain an acceptable $\chi^2/d.o.f.$ for solution III. In all cases
the value of  $\delta'^0_0$ will be kept fixed to the central value
when calculating uncertainties. This means that although the derivative is continuous, the uncertainties on the derivative at that point might have a small kink. Otherwise the error band becomes unrealistically large.

Concerning the elasticity function, it will be fitted through an exponential function with a negative exponent, to ensure $0\leq \eta^0_0 \leq1$. We have found that in the case of the $S$-wave, Chebyshev polynomials in this exponent produce unwanted oscillations. Thus we will use a simple phenomenological expansion in powers of $Q(s)\equiv q(s)/q_m-1$, where $q_m=q(s_m)$. In practice we have found that five terms are needed at most to obtain a good fit. Explicitly:
\begin{equation}
\eta^0_0(s)=\exp\left[-\left(\sum_{k=0}^{4} \epsilon_k Q(s)^k\right)^2\right].
\end{equation}
Continuity at the matching point fixes
\begin{equation}
\epsilon_0=\sqrt{-\log(\eta^0_0(s_m))},
\label{eq:eta00e0}
\end{equation}
and then the continuity of the derivative fixes
\begin{equation}
\epsilon_1=-\frac{4q_m^2}{\epsilon_0}\frac{{\eta^0_0}'(s_m)}{{\eta^0_0}(s_m)}.
\end{equation}
In practice only three free parameters will be needed at most to obtain an acceptable $\chi^2/d.o.f.$, and just one will be enough for solution I.
Note that the logarithm in Eq.~\eqref{eq:eta00e0} appears in the constants needed for the smooth matching, but it does not introduce any spurious analytic structure. 
As with the phase, now  $\eta^{0\,\prime}_0$ will be kept fixed to its central value
when calculating uncertainties. 

In summary, the S0-wave high-energy parameterization has six free parameters at most, but we will see that when fitting data solution II needs only four and solution I just two, setting to zero the remaining ones. Recall that up to 1.4 GeV all three solutions are compatible among themselves.

\subsection{P-wave parameterization}
\label{sec:Pparameterization}

The $\pi\pi$-scattering $P$-wave is completely dominated by the $\rho(770)$ meson, which is customarily described using simple resonance models, like variations of Breit-Wigner parameterizations. In many cases, this is fair enough.
However, even though the $\rho(770)$ is usually considered as the prototype of narrow resonance, its width is relatively large compared to its mass, 
which explains that the $\rho$-meson shape cannot be fully described with precision using a simple Breit-Wigner function or within an Isobar Model, but requires additional shape parameters~\cite{Pisut:1968zza,Lafferty:1993sx}.
Let us also recall that the  $\rho(770)$ is the main player of vector meson dominance. Actually, it saturates the most common hadronic observables, like, for instance, the hadronic total cross section $\sigma(e^+e^-\to \textrm{hadrons})$, which implies applications well beyond low-energy meson physics. 
Thus, given its relevance for Hadron Physics, we will provide  in this section an analytic parameterization to describe the $\pi\pi$ vector-isovector channel up to approximately 2 GeV.

This wave is much simpler than the S0, since the inelasticity sets in at much higher energies and is much smaller than for the S0 wave. Actually, there is no need to factorize explicitly any resonance pole in the inelastic region as we did for the \fzero, but for convenience we will use a similar analytic formalism to introduce the small inelasticity above $K\bar K$ threshold. This will allow us to continue analytically our partial wave to the complex plane and mimic the dispersive results
of \cite{GarciaMartin:2011cn} within the Lehmann ellipse. Once more, we will separate the energy regions below and above 1.4 \gev, because the latter is not tested against the dispersive representation and has inconsistent data sets that will be fitted separately later.

\subsubsection{P-wave parameterization below 1.4 \gev}

Thus, as we did for the S0 wave, we build our partial wave as $S^1_1=S^1_{1,\text{conf}}S^1_{1,in}$, which for the partial-wave amplitudes implies
\begin{equation}
t^1_1(s)=t^1_{1,\text{conf}}(s)+t^1_{1,in}(s)+2i\sigma(s)t^1_{1,\text{conf}}(s)\,t^1_{1,in}(s).
\label{eq:finalP}
\end{equation}
The elastic region is dominated by the $\rho(770)$ resonance and its peak mass will be imposed with an explicit factor in a purely elastic contribution, $t^1_{1,\text{conf}}(s)$, parameterized with a conformal expansion as follows:
\begin{align}
&t^1_{1,\text{conf}}(s)=
\frac{1}{\sigma(s)}\frac{1}{\Phi^1_1(s)-i}, \quad s<1.4 \gev\\
&\Phi^1_1(s)=\frac{\sqrt{s}}{2q^3(s)}(m_\rho^2-s)\left(\frac{2 m_\pi^3}{m_\rho^2 \sqrt{s}}+\sum_{n=0}^N{B_n \omega(s)^n}\right).
\label{eq:generalconformalP}
\end{align} 
Once again, $\Phi^1_1(s)=\cot\delta^1_1(s)$ in the elastic region $s\leq 4m_K^2$.
As with the S0 wave, the term $\sim1/\sqrt{s}$ within the parenthesis removes spurious ghosts but makes an almost irrelevant contribution to the fit.
For this wave it will be enough to set $N=4$  to obtain a good overall $\chi^2/d.o.f$ in this region. As before, the conformal variable is  defined as
\begin{equation}
    \omega(s)=\frac{\sqrt{s}-\alpha \sqrt{s_0-s}}{\sqrt{s}+\alpha \sqrt{s_0-s}},
    \label{eq:coformalvariableP}
\end{equation}
but now $s_0=\left(1.43 \gev\right)^2$ and in order to get an error band whose shape is closer to the actual spread of data, $\alpha$ is chosen so that the expansion center is near the $\pi\pi$ threshold. Values ranging from 0.2 to 0.5 make a suitable parameterization and we use $\alpha=0.3$.

We have already commented that, in contrast to the large \fzero effects in the S0 wave, for the $P$-wave inelastic effects are very tiny below 1.12 \gev and very small below $\sqrt{s_m}\equiv 1.4 \gev$. Actually, if one is not interested in very high accuracy, using the conformal part of the parameterization alone is almost indistinguishable from using our full partial-wave below 1.12 GeV. However, this very small inelasticity is relevant for the accurate fulfillment of dispersion relations (particularly the  $\pi^+\pi^0$ Forward Dispersion Relation, see below).
Hence, we will include an inelasticity right from $K\bar K$ threshold by means of another $t^1_{1, in }(s)$ amplitude defined as
\begin{eqnarray}
\label{eq:ineP}
t^1_{1,in}(s)&=&\frac{e^{i \delta^1_{1,in}(s)}-1}{2 i\sigma(s)}, \\
\delta^1_{1,in}(s)&=&\bar{J}(s,m_K) \left(K_0+K_1 \frac{s}{m_K^2}\right) \frac{q_\pi^3(s)}{\sqrt{s} \, m_\pi^2} \frac{q_K^2(s)}{m_K^2}, \nonumber
\end{eqnarray}
where $q_i=\sqrt{s/4-m_i^2}$ are the CM momenta of the two-pion or two-kaon system
and the analytic function $\bar J(s,m_i)$ was defined in Eq.\eqref{eq:Jbar}. Thus, the whole $t^1_1$ amplitude in Eq.\eqref{eq:finalP} has the appropriate kinematic behavior. Namely, the elasticity behaves as $q_K^3(s)/\sqrt{s}$ near $K \bar K$ threshold, whereas the phase shift behaves as $q_\pi^3(s)/\sqrt{s}$ near the $\pi \pi$ one.
In addition, since $\bar J(s,m_K)$ is real below $K\bar K$ threshold, this ensures that the whole $t^1_1$ is elastic for $s<4 m_K^2$ (no inelasticity has been observed there). 
Above $K\bar K$ threshold $\bar J(s,m_K)$ has an imaginary part and therefore both $t^1_{1, in}$ and the whole $t^1_1$ become inelastic. The advantage of using the $\bar J$ function is that this parameterization is analytic in the whole energy region from $s=0$ up to 1.4 GeV
and provides a straightforward analytic continuation to the complex plane that the usual step functions do not provide. Thus we can also continue $t^1_1$ to the complex plane within its Lehmann ellipse. Note also that $t^1_{1, in}$ contributes with a tiny phase shift below the $K \bar K$ threshold, which is on average more than two orders of magnitude smaller than the one coming from the conformal mapping. Thus, as we commented above, in the elastic region $t^1_{1, \text{conf}}$ by itself alone gives a remarkably good description of the whole partial wave. However, in order to get to 1.4 GeV the full Eq.\eqref{eq:finalP} is needed.

In practice we have found that just two constants $K_0, K_1$  together with the conformal parameterization in Eq.~\eqref{eq:generalconformalP} are enough to describe the phase shift and inelasticity 
in the real axis below 1.4 \gev as well as the complex plane of the P-wave 
within the Lehmann ellipse, including the $\rho(770)$ pole obtained in the \cite{GarciaMartin:2011cn} dispersive analysis.

\subsubsection{P-wave parameterization above 1.4 \gev}

As before with the S0 wave, above $\sqrt{s_m}\equiv 1.4 \gev$ we will provide just phenomenological fits to the $P$-wave data, ensuring a continuous matching for the phase and elasticity as well as their derivatives. The matching procedure is similar to that for the S0 wave.

For the $P$-wave the phase shift will be described using Chebyshev polynomials again. Once the matching with the previous parameterization below 1.4 GeV is implemented, we can obtain $\delta^1_1(s_m)$ and  $\delta^{\prime\,1}_1(s_m)$, so that the phase shift in the region  above 1.4 GeV reads:
\begin{align}
    \delta^1_1(s)&=\delta^1_1(s_m)+\Delta^1_1 [x_1(\omega_2(s))+1]+d_0 [x_2(\omega_2(s))-1] \nonumber \\
    &+d_1 [x_3(\omega_2(s))+1], \nonumber \\
    \Delta^1_1&=\frac{\delta^{\prime\,1}_1(s_m)}{\omega'_2(s_m)}-d_0 x'_2(-1) -d_1 x'_3(-1) \nonumber \\
    & = \frac{\delta^{\prime\,1}_1(s_m)}{\omega'_2(s_m)}+4d_0-9d_1.\label{eq:highphase11} 
\end{align}
We have considered up to three Chebyshev polynomials because, when fitting the data on the P-wave phase above 1.4 \gev in the next section, we will just need two degrees of freedom, $d_0$ and $d_1$, to obtain an acceptable $\chi^2/d.o.f$ for all solutions. The polynomials variable is the same as in the S0 case:
\begin{align}
    \omega_2(s)&=2\frac{\sqrt{s}-\sqrt{s_m}}{2 \gev-\sqrt{s_m}}-1. \nonumber \\
\end{align}
Also, as it happened in the S0 case, for the calculation of uncertainties we will keep $\delta'^1_1(s_m)$ fixed to its central value.

As we did for the S-wave above 1.4 GeV, for the P-wave elasticity we will use  an exponential with a negative exponent to ensure $0\leq \eta^1_1\leq1$. This time we have found that using up to the fourth Chebyshev polynomial is good enough to describe this exponent in all cases and do not produce unwanted oscillations. Thus we write:
\begin{equation}
    \eta^1_1(s)=\exp\left[-\left(\epsilon_0+\sum_{k=1}^4 \epsilon_k \Big(x_k(\omega_2(s))-(-1)^k\Big)\right)^2\right].
\end{equation}
Once again, continuity at the matching point fixes
\begin{equation}
    \epsilon_0=\sqrt{-\log(\eta^1_1(s_m))},
    \label{eq:eta11e0}
\end{equation}
whereas the continuity of the derivative imposes
\begin{align}
    \epsilon_1&=-\frac{\eta^{\prime\,1}_1(s_m)}{2 \epsilon_0 \eta^1_1(s_m)\omega'_2(s_m)}-\epsilon_2 x'_2(-1)\nonumber\\ 
              &\quad-\epsilon_3 x'_3(-1)-\epsilon_4 x'_4(-1)\nonumber\\
              &=-\frac{\eta^{\prime\,1}_1(s_m)}{2 \epsilon_0 \eta^1_1(s_m)\omega'_2(s_m)}+4\epsilon_2-9\epsilon_3+16\epsilon_4.
\end{align}
Thus, in practice, three free parameters are needed at most.
Once again we remark that the logarithm in Eq.~\eqref{eq:eta11e0} appears in the constants needed for the smooth matching but it does not introduce any spurious analytic structure. 
For the calculation of uncertainties we will keep ${\eta'^1_1}(s_m)$ fixed to its central value. Thus the central value of the derivative is continuous but its uncertainties might show a small kink.

\section{Determination of Parameters}

The aim of this work is to provide a relatively simple global description for each one of the S0 and P waves of $\pi\pi\rightarrow \pi\pi$ scattering, incorporating all analytic constrains at low energies, including Adler zeros, while also describing the existing data up to 2 GeV. They should also be consistent with the dispersive analysis of data up to 1.4 GeV in~\cite{GarciaMartin:2011cn}. Moreover, such parameterizations should provide 
also simple but realistic estimates of the uncertainties.
In the previous section we have provided such simple and ready to use parameterizations. In this sections we will determine the value of their parameters.

Let us recall that the CFD parameterizations of \pipi scattering partial waves obtained in~\cite{GarciaMartin:2011cn} are data fits constrained to fulfill a group of forward dispersion relations up to 1.43 \gev, together with the more sophisticated Roy and GKPY equations for the partial waves, applicable up to roughly 1.1 \gev. However they were parameterized with piecewise functions and we now want to mimic them and their uncertainties with a global parameterization. Thus, in the real axis for $4m_\pi^2<s<s_m$ the CFD partial waves  of~\cite{GarciaMartin:2011cn} will be fitted.  The CFD has much smaller uncertainties than the output of the dispersion relations themselves, and this is why it is preferred to build a more accurate result. We will impose just the phase shift up to the inelastic $K \bar K$ threshold, and both the phase shift and elasticity above it.

In addition, we want our new parameterization to be consistent with the dispersive result in the subthreshold region and in the complex plane, particularly with the resonance pole positions and residues. The CFD are piecewise functions and although some of the pieces contain fair approximations to the poles, they do not provide accurate results in the complex plane. Therefore below the elastic threshold, and in the complex plane, we will fit our global parameterization to the output of GKPY equations, which produces narrower errors than that of Roy equations, while both are compatible among themselves in the whole complex plane and real axis. The fit will run from about $\re s \sim (0 \gev)^2$ to $\re s \sim (1.12 \gev)^2$, but always inside the applicability region of the GKPY or Roy dispersion relations, which can be found in~\cite{Roy:1971tc,Ananthanarayan:2000ht,Caprini:2005zr}. Using such a vast region we are able to describe the scattering lengths, the Adler zeros in the S0 wave, and the \sig and $\rho(770)$ pole positions and couplings (the \fzero is fixed as input). Due to the smaller uncertainties in the real axis, the final errors bars of our parameterization in the complex plane are smaller than the dispersive ones. 

All these features will be imposed on our parameterization by means of a $\chi^2/d.o.f.$ function,  over a grid of points separated by 10 MeV both in the real and imaginary directions within the GKPY/Roy equations applicability region. The input values and uncertainties in this $\chi^2$ are those of the CFD in the physical region below $s_m$ and of the GKPY output \cite{GarciaMartin:2011cn} in the subthreshold region and in the rest of the complex plane.
 Nevertheless, the statistical meaning of the $\chi^2/d.o.f.\sim 1$ loses part of its purpose, as the results coming from dispersion relations are smooth functions instead of normally distributed points, and their uncertainties are totally correlated between bins. As a result, a value lower than 1 is frequently expected, and we will consider all results below or around 1 as good descriptions of our dispersion relations.
 
 Finally, let us recall that  above 1.43 \gev  no dispersive result exists, thus we will make use of the available experimental data. As explained in section~\ref{sec:input} the data sources in this energy region produce three different plausible solutions: the first one, called solution I in this work will fit data from~\cite{Hyams:1973zf, Grayer:1974cr,Becker:1978ks,Kaminski:1996da}. The second one, that we will call solution II, fits data from a later reanalysis by the CERN-Munich collaboration~\cite{Hyams:1975mc}. Finally, solution III uses a recent update \cite{Ochs:2013gi} of the (- + -) data solution in~\cite{Hyams:1975mc}.

\subsection{$S0$-wave fit}

\begin{figure}
\centering
\includegraphics[width=0.45\textwidth]{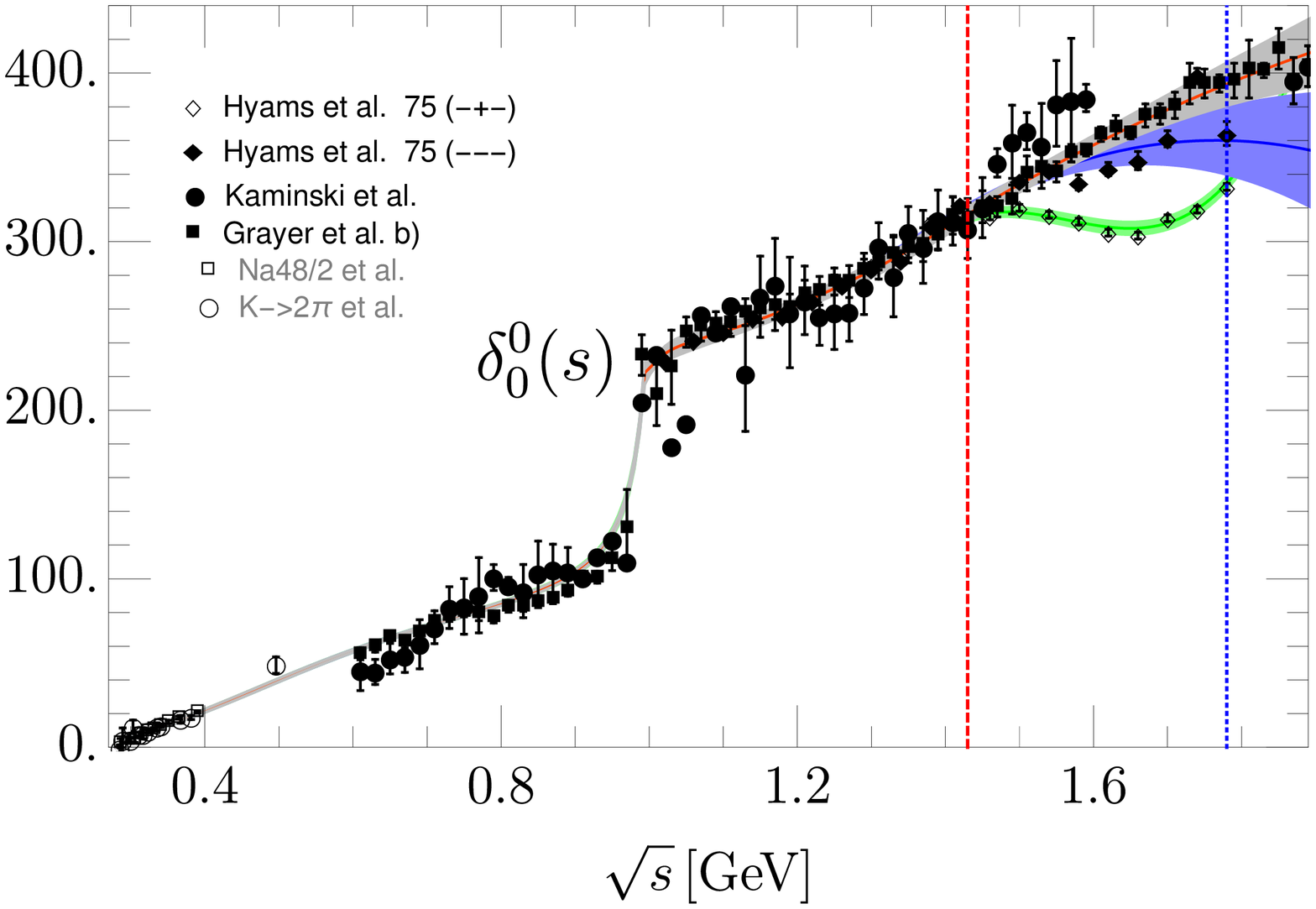}\vspace{.1cm} \\
\includegraphics[width=0.45\textwidth]{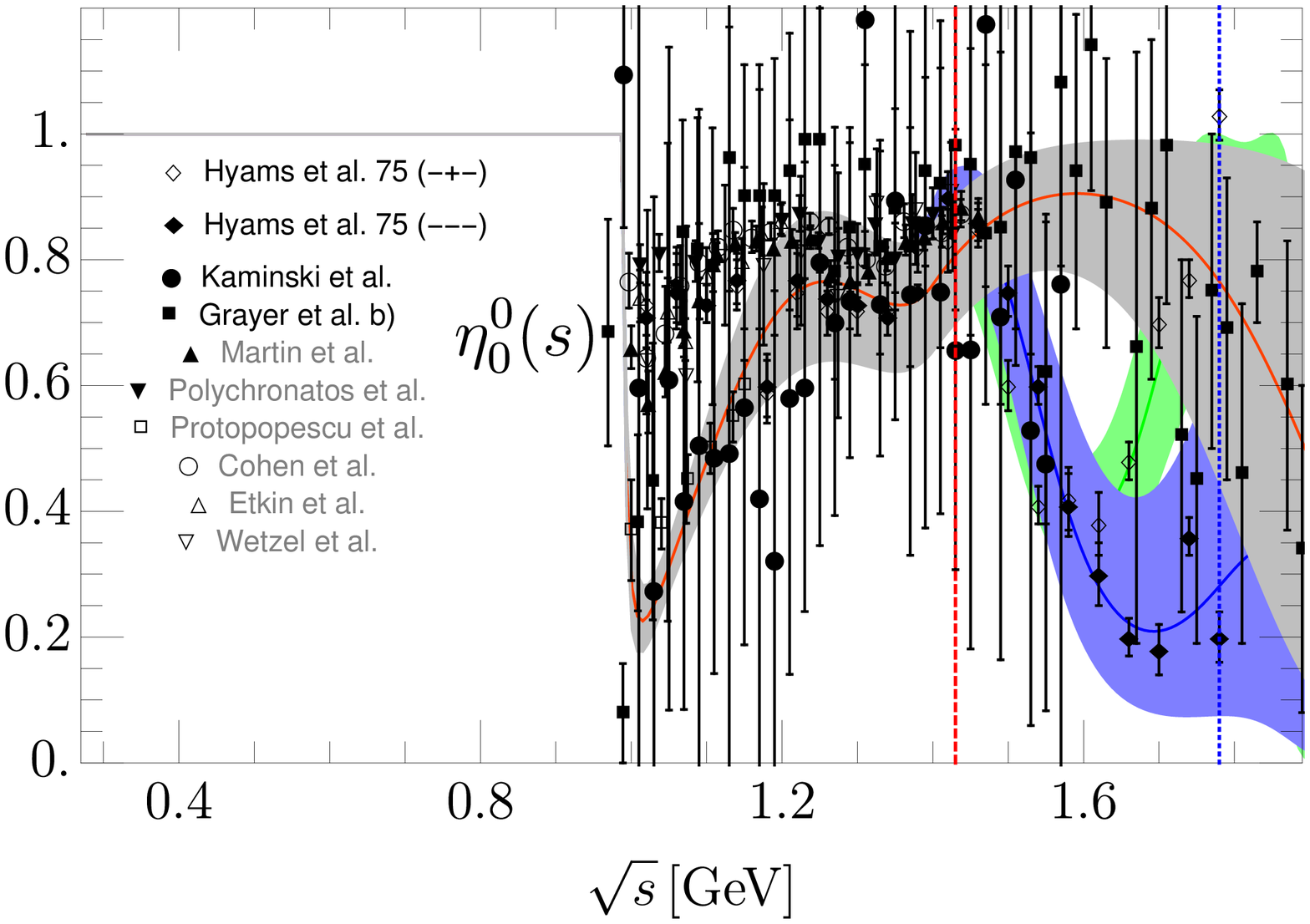}
\caption{ \small \label{fig:shift-inelas} Comparison of solutions I, II and III (Tables \ref{tab:paraI}, \ref{tab:paraII}, \ref{tab:paraIII}) versus data. 
The gray, blue and green bands correspond to the uncertainty of solutions I, II and III, respectively. 
Above 1.4 GeV, solution I fits the data of~\cite{Becker:1978ks,Kaminski:1996da} (solid circles) and~\cite{Hyams:1973zf,Grayer:1974cr} (solid squares),  solution II fits~\cite{Hyams:1975mc} (solid diamonds) and solution III fits the updated (- + -) data from \cite{Ochs:2013gi} (hollow diamonds). The data coming from~\cite{Batley:2010zza} (empty squares) and~\cite{Cohen:1980cq} (empty circles) for the phase shift and~\cite{Martin:1979gm} (solid triangle up), ~\cite{Polychronakos:1978ur}(solid triangle down),~\cite{Protopopescu:1973sh} (empty squares),~\cite{Cohen:1980cq} (empty circles),~\cite{Etkin:1981sg} (empty triangle up) and~\cite{Wetzel:1976gw} (empty triangle down) for the elasticity are just shown for comparison.
The red-dashed vertical line separates the region where the fits describe both data and dispersion relation results, from the region above, where the parameterization is just fitted to data. The blue-dotted vertical line stands at the energy of the last data point of solutions II and III.}
\end{figure}

We show in Fig.~\ref{fig:shift-inelas} solutions I, II and III of our new S0-wave parameterization up to 1.9 GeV. 
Their parameters are listed in Table~\ref{tab:paraI} for solution I,  Table~\ref{tab:paraII} for solution II and Table~\ref{tab:paraIII} for solution III. By construction, they are almost identical up to 1.4 GeV. Nevertheless, there is an almost imperceptible deviation between them in the inelastic region below 1.4 \gev due to their matching to different solutions above 1.4 GeV. Actually, above this splitting point the solutions are fairly different either on their phase, elasticity or both.

It is worth noticing that the uncertainties of solution II are larger for the phase shift, due to the scarcity of data above 1.5 \gev. In addition,  the elasticity data forces solution II to drop first and then to raise in the region between 1.5 and 1.8 GeV, which is hard to explain in terms of known resonances. 
Above 1.8 GeV there are no data for this solution and our functional forms would give even more oscillations, for which there is no evidence. Thus, to avoid further oscillatory behavior above 1.8 GeV, we have included four elasticity data points above that energy coming from~\cite{Hyams:1973zf, Grayer:1974cr} which have huge uncertainties but stabilize the fit.  In contrast solution I slowly becomes more and more inelastic as the energy increases which is more natural if more and more channels are open. The phase motion of solution III and the relatively sharp dip of the elasticity, which are quite different from the other solutions, are hints of the presence of the $f_0(1500)$ resonance.

\begin{table}[!hbt] 
\caption{ \small Fit parameters of the global parameterization for the $S0$-wave solution I. $s_p$ is the \fzero pole position from the dispersive analysis in \cite{GarciaMartin:2011jx}.}
\centering 
\begin{tabular}{c c | c c | c c} 
\hline\hline
\multicolumn{2}{c}{$t^0_{0,\text{conf}}$} & \multicolumn{2}{c}{$t^0_{f_0}$} & \multicolumn{2}{c}{$\sqrt{s}>1.4 \gev$} \\
\hline\hline  
\rule[-0.05cm]{0cm}{.35cm}$B_0$ & 12.2$\pm$0.3 & $K_0$ &   5.25$\pm$0.28 & $d_0$ &   -5.4$\pm$3.7\\
\rule[-0.05cm]{0cm}{.35cm}$B_1$ & -0.9$\pm$1.1 & $K_1$ &  -4.40$\pm$0.16 &  $d_1$ & $\equiv0$  \\ 
\rule[-0.05cm]{0cm}{.35cm}$B_2$ & 15.9$\pm$2.7  & $K_2$ &     0.175$\pm$0.155 & $d_2$ & $\equiv0$\\ 
\rule[-0.05cm]{0cm}{.35cm}$B_3$ & -5.7$\pm$3.1 & $K_3$ &   -0.28$\pm$0.06 &  $\epsilon_2$ &   10.3$\pm$4.0 \\ 
\rule[-0.05cm]{0cm}{.35cm}$B_4$ & -22.5$\pm$3.7 & &  & $\epsilon_3$ & $\equiv0$\\ 
\rule[-0.05cm]{0cm}{.35cm}$B_5$ &  6.9$\pm$4.8 & $\re \sqrt{s_p}$  & 0.996$\pm$7 \gev& $\epsilon_4$ & $\equiv0$\\ 
\rule[-0.05cm]{0cm}{.35cm}$z_0$ &  0.137$\pm$0.028 \gev& $\im \sqrt{s_p}$  & -0.025$\pm$8 \gev& &\\ 
\hline\hline
\end{tabular} 
\label{tab:paraI} 
\end{table}

\begin{table}[!hbt] 
\caption{ \small Fit parameters of the global parameterization for the $S0$-wave solution II. $s_p$ is the \fzero pole position from the dispersive analysis in \cite{GarciaMartin:2011jx}.}
\centering 
\begin{tabular}{c c | c c | c c} 
\hline\hline
\multicolumn{2}{c}{$t^0_{0,\text{conf}}$} & \multicolumn{2}{c}{$t^0_{f_0}$} & \multicolumn{2}{c}{$\sqrt{s}>1.4 \gev$} \\
\hline\hline  
\rule[-0.05cm]{0cm}{.35cm}$B_0$ & 12.2$\pm$0.3 & $K_0$ &   4.97$\pm$0.08 & $d_0$ &   -16.5$\pm$6.2\\
\rule[-0.05cm]{0cm}{.35cm}$B_1$ & -1.2$\pm$0.8 & $K_1$ &  -4.72$\pm$0.08 & $d_1$ & $\equiv0$  \\ 
\rule[-0.05cm]{0cm}{.35cm}$B_2$ & 15.5$\pm$1.5  & $K_2$ &  -0.04$\pm$0.18  & $d_2$ & $\equiv0$\\ 
\rule[-0.05cm]{0cm}{.35cm}$B_3$ & -6.0$\pm$1.5 & $K_3$ &   -0.31$\pm$0.04 & $\epsilon_2$ &  160.8$\pm$2.4\\ 
\rule[-0.05cm]{0cm}{.35cm}$B_4$ & -21.4$\pm$1.3 & & & $\epsilon_3$ & -715.5$\pm$8.5   \\ 
\rule[-0.05cm]{0cm}{.35cm}$B_5$ &  6.3$\pm$4.5 & $\re \sqrt{s_p}$ & 0.996$\pm$7 \gev& $\epsilon_4$ &   -937.3$\pm$25.0\\ 
\rule[-0.05cm]{0cm}{.35cm}$z_0$ &  0.135$\pm$0.031 \gev& $\im \sqrt{s_p}$  & -0.025$\pm$8 \gev& &\\ 
\hline\hline
\end{tabular} 
\label{tab:paraII} 
\end{table}

\begin{table}[!hbt] 
\caption{ \small Fit parameters of the global parameterization for the $S0$-wave solution III. $s_p$ is the \fzero pole position from the dispersive analysis in \cite{GarciaMartin:2011jx}.}
\centering 
\begin{tabular}{c c | c c | c c} 
\hline\hline
\multicolumn{2}{c}{$t^0_{0,\text{conf}}$} & \multicolumn{2}{c}{$t^0_{f_0}$} & \multicolumn{2}{c}{$\sqrt{s}>1.4 \gev$} \\
\hline\hline  
\rule[-0.05cm]{0cm}{.35cm}$B_0$ & 12.3$\pm$0.3 & $K_0$ &   5.26$\pm$0.08 & $d_0$ &   73.4$\pm$1.5\\
\rule[-0.05cm]{0cm}{.35cm}$B_1$ & -1.0$\pm$0.9 & $K_1$ &  -4.64$\pm$0.04 & $d_1$ & 27.3$\pm$0.4  \\ 
\rule[-0.05cm]{0cm}{.35cm}$B_2$ & 15.7$\pm$1.7  & $K_2$ &  0.10$\pm$0.07  & $d_2$ & -0.3$\pm$0.2\\ 
\rule[-0.05cm]{0cm}{.35cm}$B_3$ & -6.0$\pm$1.6 & $K_3$ &   -0.29$\pm$0.04 & $\epsilon_2$ &  171.6$\pm$2.0\\ 
\rule[-0.05cm]{0cm}{.35cm}$B_4$ & -22.1$\pm$1.2 & & & $\epsilon_3$ & -1038.8$\pm$8.3   \\ 
\rule[-0.05cm]{0cm}{.35cm}$B_5$ &  7.1$\pm$2.8 & $\re \sqrt{s_p}$ & 0.996$\pm$7 \gev& $\epsilon_4$ &   1704.7$\pm$30.8\\ 
\rule[-0.05cm]{0cm}{.35cm}$z_0$ &  0.136$\pm$0.035 \gev& $\im \sqrt{s_p}$  & -0.025$\pm$8 \gev& &\\ 
\hline\hline
\end{tabular} 
\label{tab:paraIII} 
\end{table}

\begin{figure}
\centering
\includegraphics[width=0.5\textwidth]{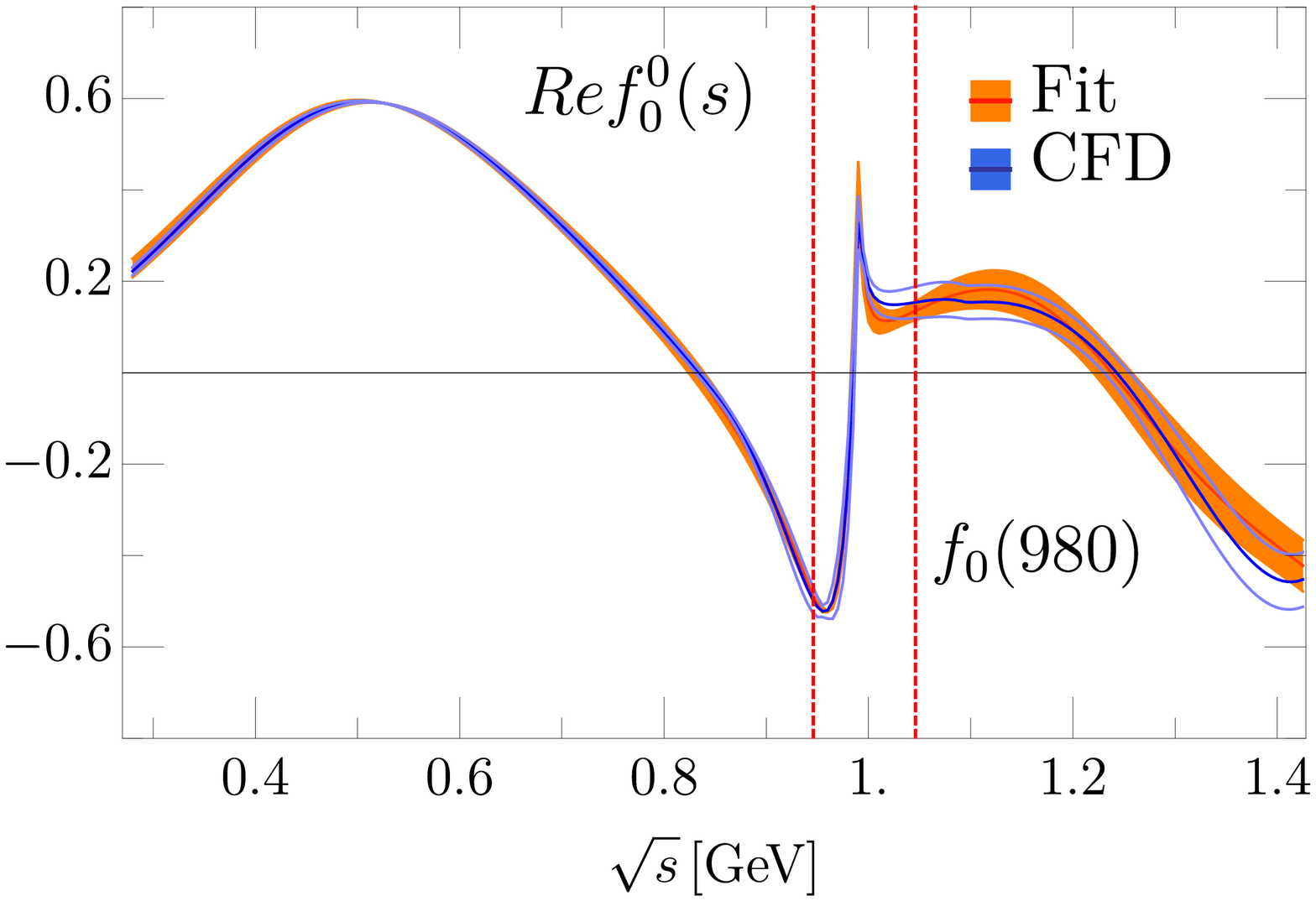}\vspace{.3cm} \\
\includegraphics[width=0.5\textwidth]{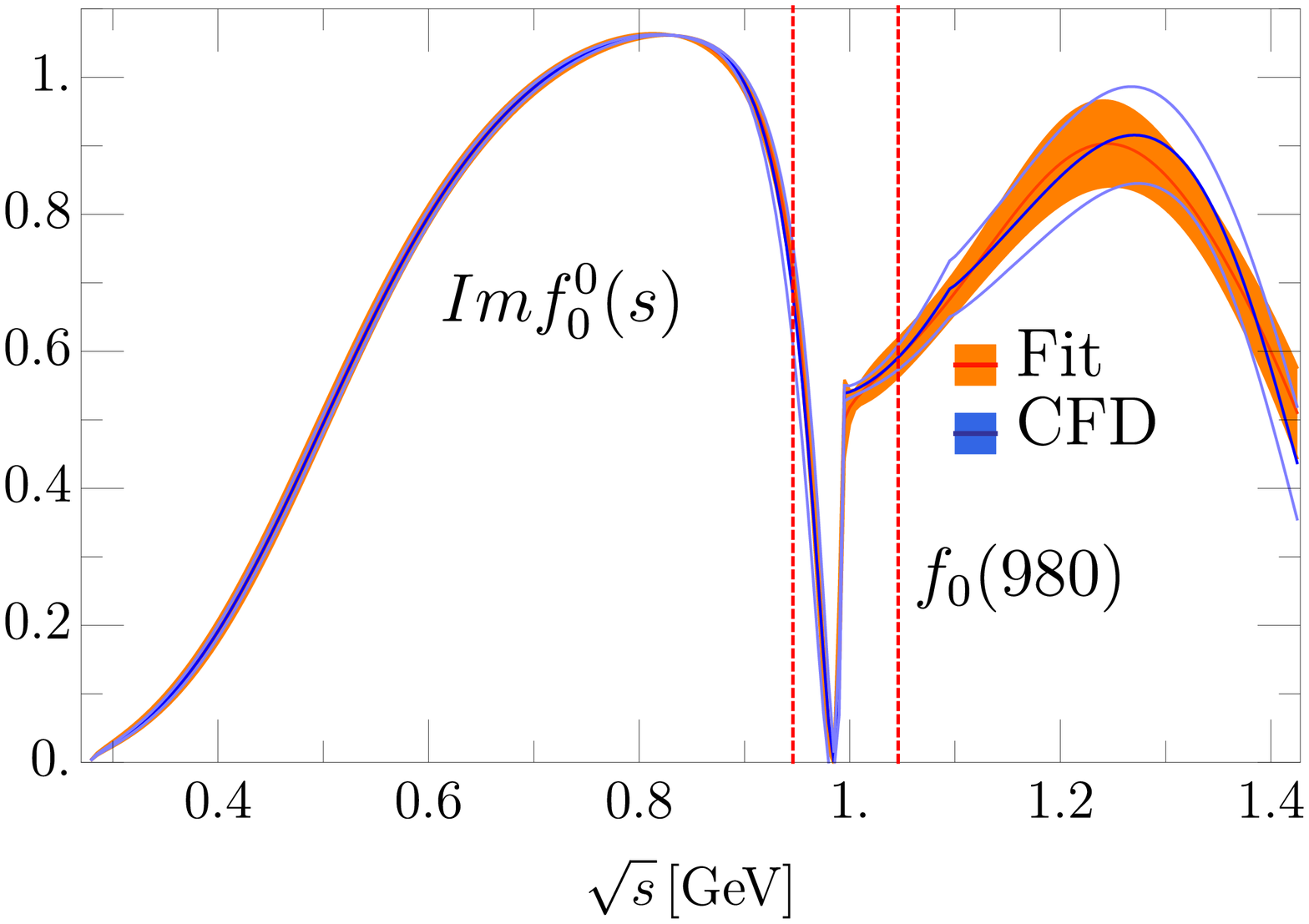}
\caption{ \small  \label{fig:real-imag} Comparison between the CFD fit in~\cite{GarciaMartin:2011cn} (blue) and solution I (Table \ref{tab:paraI}, orange band). The energy region dominated by the \fzero pole is delimited between the red dashed lines.}
\end{figure}

Concerning the compatibility with the dispersive results in~\cite{GarciaMartin:2011cn}, we show in Fig.~\ref{fig:real-imag} the comparison between the CFD analysis of~\cite{GarciaMartin:2011cn} and our solution I. 
Up to 1.4 GeV it is enough to refer to solution I as the global solution, because it is the simplest and all them are almost indistinguishable below 1.4 GeV. The relevant observation from Fig.~\ref{fig:real-imag} is that the piecewise CFD and our new parameterization look almost the same below the $K \bar K$ threshold and are also very similar and compatible above it. The sharp structure in the  region between the two vertical lines in Fig.~\ref{fig:real-imag} is dominated by the \fzero contribution that we have factored out explicitly in our global parameterization.

All in all, this new parameterization is consistent with the GKPY dispersive data analysis, its output in the complex plane, as well as with the threshold parameters, the Adler zero, the positions of both \sig and \fzero poles, and the inelastic region up to 1.43 \gev, which was consistent with Forward Dispersion Relations. 
This consistency is illustrated in Table~\ref{tab:chi2} where we show the $\chi^2/d.o.f.\equiv\hat \chi^2$ of our fit with the new parameterization in different regions: $\hat \chi_1^2$ from $\pi\pi$ to $K\bar K$ threshold, $\hat \chi_2^2$ from
$K\bar K$ threshold to 1.4 GeV, $\hat \chi_{\IC}^2$ in the complex plane within the applicability region, $\hat\chi^2_{\delta}$ for the phase above 1.4 GeV and  $\hat\chi^2_{\eta}$ for the elasticity above 1.4 GeV. All of them are smaller or equal to one for any of our three solutions.

Moreover, this S0-wave global parameterization up to 1.4 GeV is fully consistent with the dispersion relations described in the the GKPY dispersive analysis ~\cite{GarciaMartin:2011cn}.
Of course, for such calculation we also need the input for other partial waves and high-energy input described in ~\cite{GarciaMartin:2011cn} and thus we have relegated this discussion to appendix \ref{app:dr}.

\begin{table}[!hbt] 
\caption{ \small Results in terms of $\hat \chi^2$ of the S0 solutions I, II  and III in different regions. }
\centering 
\begin{tabular}{c | c  c  c | c c} 
\hline\hline
 & $\hat\chi^2_{1}$ & $\hat\chi^2_{2}$ & $\hat\chi^2_{\IC}$ & $\hat\chi^2_{\delta}$ & $\hat\chi^2_{\eta}$\\
\hline\hline  
\rule[-0.05cm]{0cm}{.35cm} Solution I  & 0.2 & 0.5 & 0.4 & 0.5 & 0.4\\
\rule[-0.05cm]{0cm}{.35cm} Solution II  & 0.2 & 0.4 & 0.4 & 1.0 & 1.0\\
\rule[-0.05cm]{0cm}{.35cm} Solution III  & 0.2 & 0.4 & 0.4 & 1.0 & 0.9\\
\hline\hline
\end{tabular} 
\label{tab:chi2} 
\end{table}

\subsubsection{Poles, Couplings and Low Energy Parameters}

\begin{figure}
\centering
\includegraphics[width=0.5\textwidth]{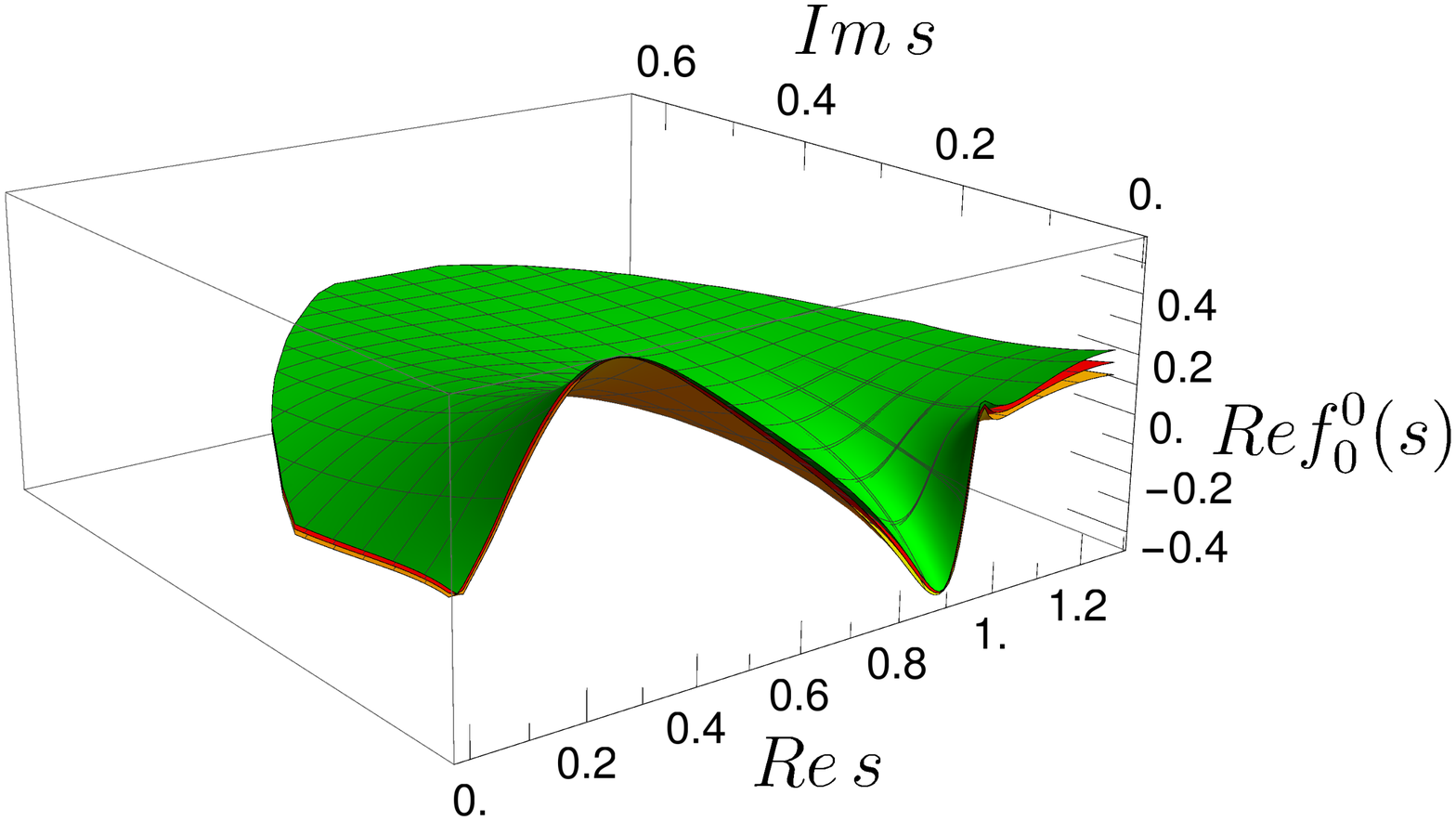} \\
\includegraphics[width=0.5\textwidth]{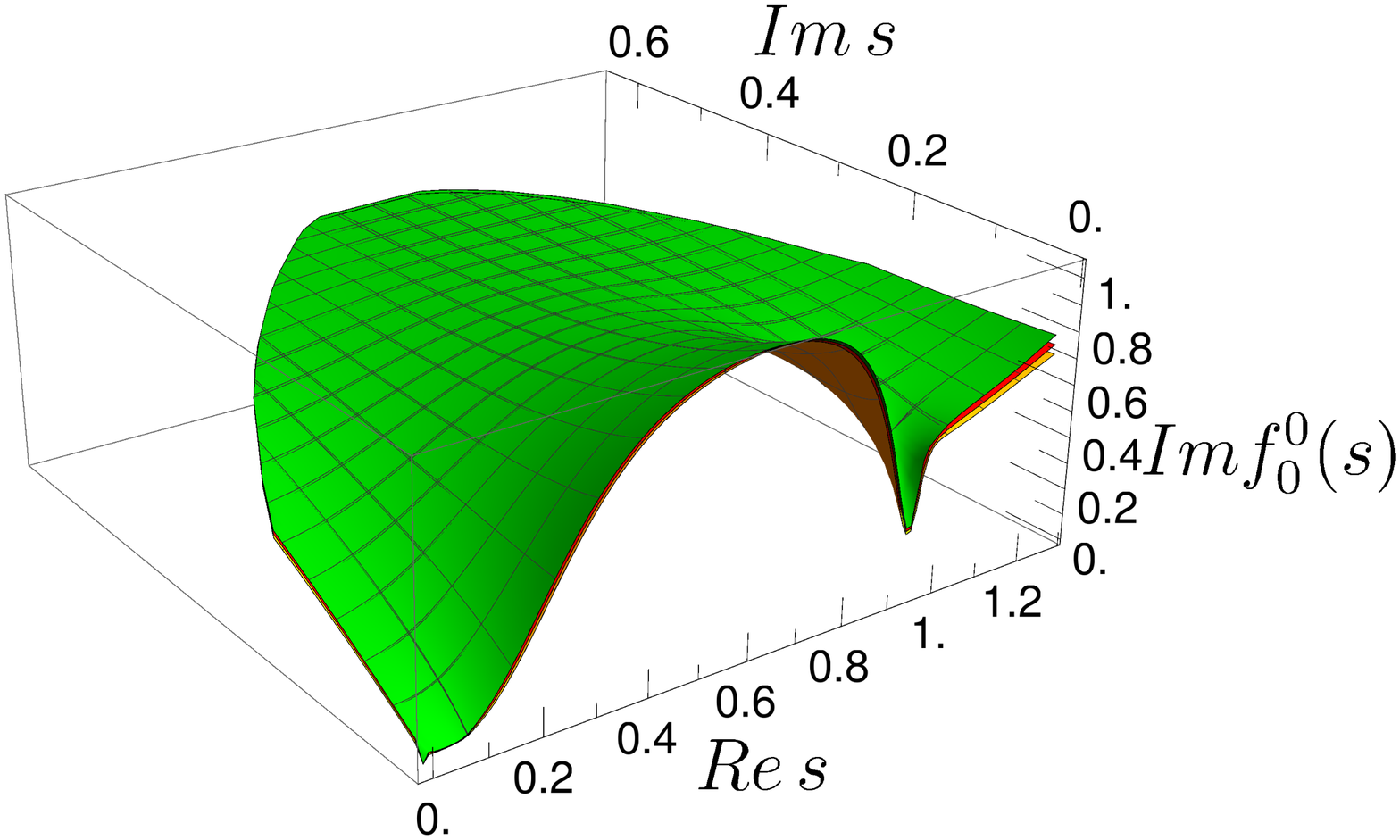}
\caption{ \small Real (top) and imaginary (bottom) parts of the scalar-isoscalar partial wave in the first Riemann sheet of the complex-$s$ plane, within the applicability region of GKPY/Roy equations. There are actually three surfaces on each plot: one for the central value, one for the upper uncertainty and another one for the lower uncertainty band. Note that the behavior of the parameterization is smooth and the uncertainties are small compared to the typical scale of the analytic structures, even deep in the complex plane. We plot solution I, since solutions I, II and III are almost identical in this region. \label{fig:complex-plane} }
\end{figure}

\begin{figure}
\centering
\hspace*{-1cm}\includegraphics[width=0.45\textwidth]{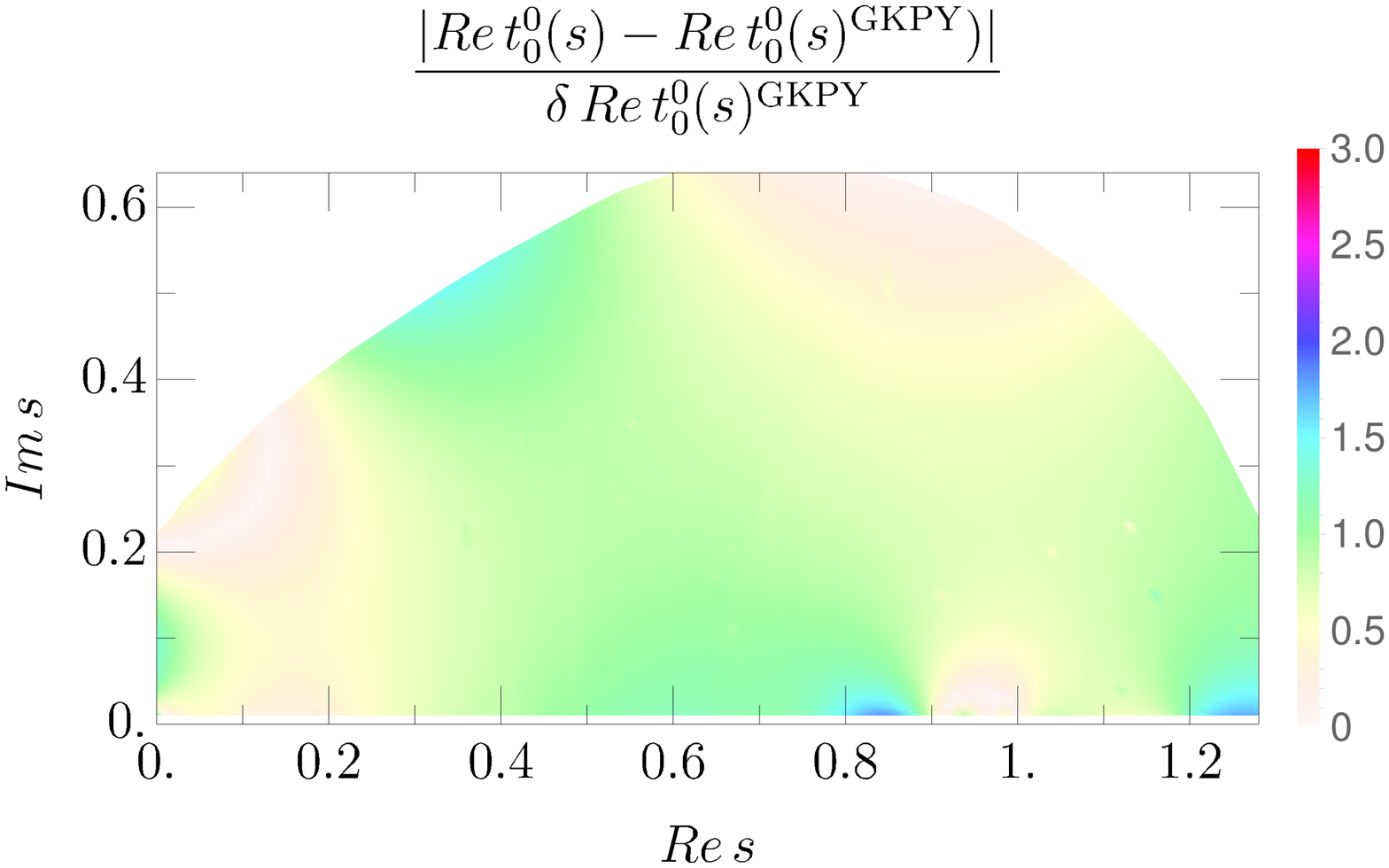}\\
\hspace*{-1cm}\includegraphics[width=0.45\textwidth]{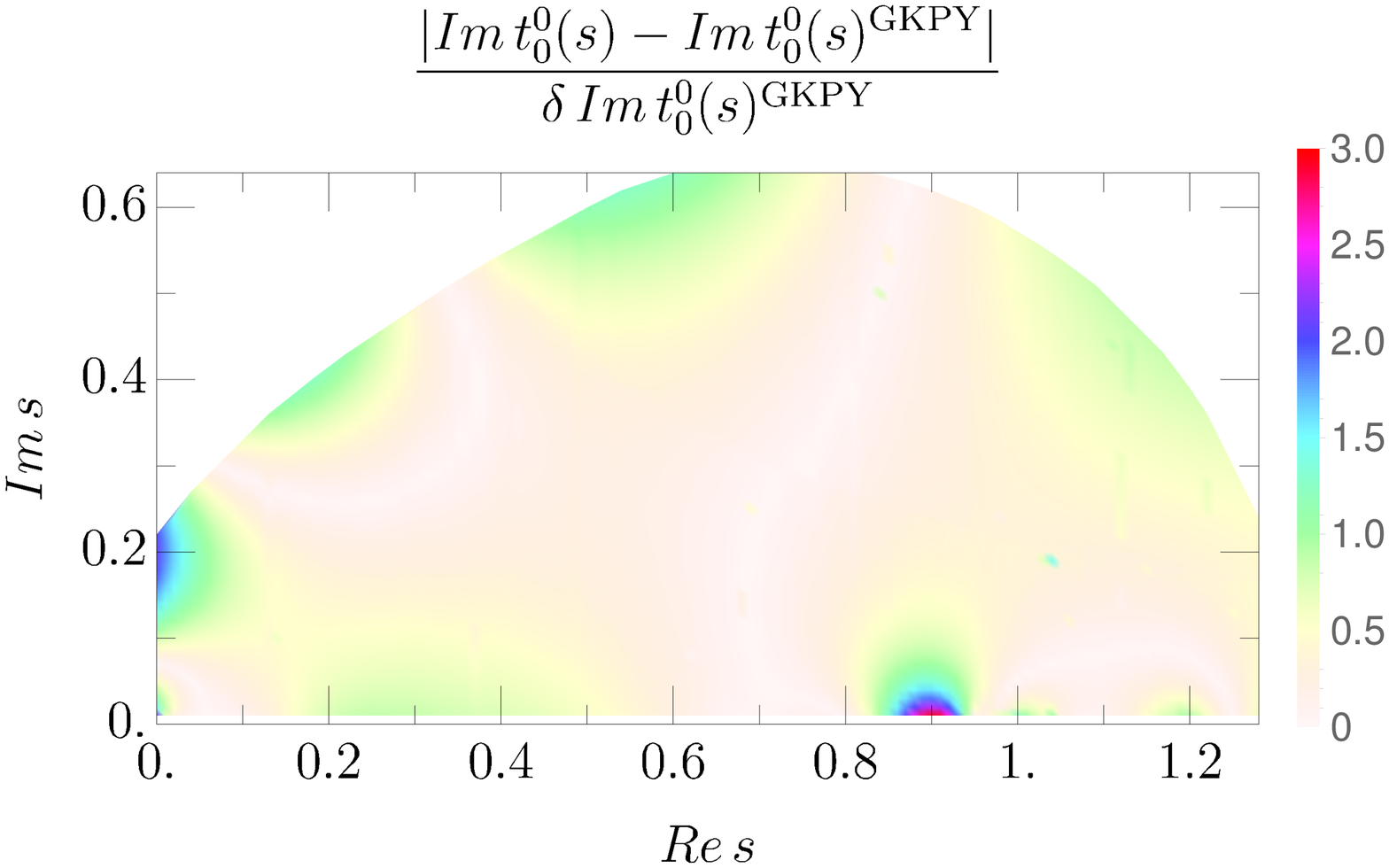}
\caption{ \small Within the applicability region of GKPY equations in the complex-$s$ plane, we show the absolute value of the differences between the real (top) and imaginary (bottom) parts of the global parameterization and the GKPY equations, divided by the uncertainty of the latter. We plot results for solution I but the other two are identical in this region.
\label{fig:complex-contour} }
\end{figure}

As explained above, the global parameterization is also constrained to describe the dispersive results in the whole complex energy-squared plane. This produces a stable and accurate description of the \sig resonance parameters. Actually, in Fig.~\ref{fig:complex-plane} we show our parameterization and its uncertainties in the first Riemann sheet of the complex plane, which reproduces the output of GKPY equations.
In order to see the consistency with the GKPY dispersive result, in the upper panel of Fig.~\ref{fig:complex-contour} we show the absolute values of the differences between the real part of our new parameterization and the GKPY result divided by the uncertainty of the latter. In the lower panel we show a similar plot for the imaginary parts.
Note that our new parameterization lies within the uncertainties of the GKPY for the most part of the region.
The only place where there are sizable differences beyond two standard deviations is for Im$\,t_0^0$ in the real axis around 0.9 GeV, but this is the matching point of the two pieces of the CFD parameterization, whereas the GKPY output is much smoother. Thus, our two inputs are slightly incompatible around that region and our new parameterization lies somewhere between both of them.

In addition,  we list in Table \ref{tab:poles} the parameters of both the  \sig and \fzero resonances compared to their GKPY dispersive values in~\cite{GarciaMartin:2011cn}. It is worth noticing that the uncertainties of the \sig resonance associated to this fit are a bit smaller than the GKPY determination~\cite{GarciaMartin:2011jx}. This is due to the fact that besides the GKPY output in the complex plane, we are fitting the CFD in the real axis, which has smaller uncertainties..

\begin{table}
\caption{ \small Pole positions and $\pi\pi$ couplings of both $f_0(500)$ and $f_0(980)$ resonances
from our global parameterization. Almost indistinguishable values would be obtained for solutions I, II and III. Note that they are very compatible with the GKPY dispersive results in~\cite{GarciaMartin:2011jx}.}
\centering 
\begin{tabular}{c | c | c} 
\hline\hline
 & $\sqrt{s_{pole}}$ (MeV) & $|g|$ ($\gev$)\\
\hline\hline  
\rule[-0.05cm]{0cm}{.35cm} $f_0(500)^{\text{GKPY}}$ & $(457^{+14}_{-13})-i (279^{+11}_{-7})$ & $3.59^{+0.11}_{-0.13}$\\
\rule[-0.05cm]{0cm}{.35cm} $f_0(500)$ & $(457\pm 10)-i (278\pm 7)$ & $3.46\pm0.07$\\
\hline 
\rule[-0.05cm]{0cm}{.35cm} $f_0(980)^{\text{GKPY}}$ & $(996\pm 7) -i (25^{+10}_{-6})$ & $2.3\pm0.2$\\
\rule[-0.05cm]{0cm}{.35cm} $f_0(980)$ & $(996\pm 7) -i (25\pm 8)$ & $2.28\pm0.14$\\
\hline\hline
\end{tabular} 
\label{tab:poles} 
\end{table}

As for the \fzero resonance, we have included in the fit the pole position obtained by means of the GKPY equations in~\cite{GarciaMartin:2011jx}. The main reason is that phenomenological fits cannot extract its accurate parameters in a very stable way (see for instance~\cite{Briceno:2017qmb}).
In particular, the CFD fit of~\cite{GarciaMartin:2011cn} does not provide an accurate estimate of its position and one has to rely on the numerical dispersive approach. However, with our new parameterization, the \fzero is no longer a problem, as both the data, the cusp effect and the pole position are factored out into a simple, yet versatile functional form. Once again, the coupling of the \fzero to $\pi \pi $ has smaller uncertainties than the GKPY determination, since the CFD partial wave, with its small uncertainty, is also fitted in the real axis to obtain our new parameterization.

\begin{table}[!hbt] 
\caption{ \small Adler zero and threshold parameters. The latter  in customary $m_\pi=1$ units. They are almost indistinguishable for solutions I, II and III.}
\centering 
\begin{tabular}{c | c | c} 
\hline\hline
 & This work & Dispersive result \cite{GarciaMartin:2011cn} \\
\hline\hline  
\rule[-0.05cm]{0cm}{.35cm} $\sqrt{s_{Adler}}$ & 96$\pm$20 \mev & 85$\pm$34 \mev \\
\rule[-0.05cm]{0cm}{.35cm} $a^0_0$ & 0.228$\pm$0.022 & 0.220$\pm$0.008 \\
\rule[-0.05cm]{0cm}{.35cm} $b^0_0$ & 0.266$\pm$0.009 & 0.278$\pm$0.005 \\
\hline\hline
\end{tabular} 
\label{tab:lowpara} 
\end{table}

Last, but not the least, the global parameterization yields relatively accurate threshold and sub-threshold parameters (like the Adler zero) compatible with those of the dispersive data analysis of the Madrid-Krakow group~\cite{GarciaMartin:2011cn} and therefore also with the dispersive analysis matched to two-loop ChPT of the Bern group~\cite{Colangelo:2001df,Ananthanarayan:2000ht}.

\subsection{$P$-wave fit}

Following the same procedure just applied to the S0 wave, in the physical region and below 1.4 $\gev$ we will 
fit our P-wave parameterization to the CFD P-wave of~\cite{GarciaMartin:2011cn}. Note that this P-wave parameterization describes data from both $\pi\pi$ scattering~\cite{Hyams:1975mc,Protopopescu:1973sh,Estabrooks:1974vu} and the pion vector form factor from~\cite{Barkov:1985ac,Amendolia:1986wj}, while fulfilling at the same time the GKPY/Roy equations up to 1.12 GeV and Forward Dispersion Relations up to 1.43 GeV.

While in recent years there have been no new $\pi\pi$-scattering experimental analyses, the pion-vector form factor data described in~\cite{GarciaMartin:2011cn} is nowadays outdated with respect to the  modern and precise measurements in~\cite{Lees:2012cj,Anastasi:2017eio}.   
These experimental data have been recently considered as input in a dispersive analysis for $e^+e^-\to\pi^+\pi^-$~\cite{Colangelo:2018mtw}, where the $\pi\pi$-scattering $P$-wave phase shift is obtained from the Roy-equation analysis in~\cite{Colangelo:2001df,Caprini:2011ky} and the $P$-wave phase-shift values at $\sqrt s=0.8$ and $1.15 \gev$ are considered as fit parameters. In this way, the  result depicted in Fig.~16 in~\cite{Colangelo:2018mtw} should be taken as the most precise and updated dispersive determination of the $\pi\pi$ scattering $P$-wave phase-shift. Nevertheless, this recent determination is compatible with the CFD result in~\cite{GarciaMartin:2011cn} within uncertainties\footnote{The maximum difference, in the region around $\sqrt{s}=0.8 \gev$, is below $1.5^\circ$.} and, for consistency, it will be still used as our input in the real axis. 

Once more, in the subthreshold region and in the complex plane we will fit the GKPY-equation dispersive results. As done for the scalar channel, we will only consider the energy region within the Lehmann ellipse, where both Roy and GKPY equations are formally valid. 
Above 1.43 $\gev$ there are no further dispersive results and hence we will only describe the available experimental data, which come from a single scattering experiment performed by the CERN-Munich Collaboration. 
In addition, in the vector case there is a relevant difference between the best solution of the original CERN-Munich result published in 1973~\cite{Hyams:1973zf} and the (- - -) solution of the 1975 collaboration reanalysis~\cite{Hyams:1975mc}. The revisited and modified (-+-) solution in~\cite{Ochs:2013gi} lies somewhat in between.

The behavior of the original P-wave result shows a large interference in the region between 1.5 and 1.8 \gev. Namely, within these 300 MeV, the phase shift changes by more than $20^\circ$ and the elasticity, starting from almost 1, decreases to less than 0.5 to return back to 1. This behavior could only be explained if the $\rho'$ and $\rho''$ resonances and the $K \bar{K}$ channel would interfere strongly, which is in contradiction with the experimental values for the width and couplings of these two resonances~\cite{Hanhart:2012wi,pdg}. Thus, the solution (- - -) of Hyams 75~\cite{Hyams:1975mc} is the one customarily used in the literature. However, we will fit three solutions for completeness, as we have done for the $S0$-wave. The original CERN-Munich result~\cite{Hyams:1973zf,Grayer:1974cr} will be called solution I, whereas the fit to the updated reanalysis of~\cite{Hyams:1975mc} will be called solution II
and the fit to the updated (- + -) solution in \cite{Ochs:2013gi} will be called solution III.

As previously done for the S0 wave we will fit our P-wave global parameterization described in Sec.~\ref{sec:Pparameterization} to 
a 10 MeV-spaced grid of GKPY output values within their applicability region in the complex plane and to the CFD parameterization in the real axis at energy points separated by 5 MeV. In addition we add the $\chi^2/d.o.f.$ of the data above 1.4 GeV, although for the phase shift of solutions I and II we have added $1^{\rm o}$ as a systematic uncertainty, since the nominal uncertainties in some regions are unrealistically small, particularly for solution II. 
The fit minimizes a $\chi^2/d.o.f.$ function whose uncertainties are those of the GKPY or the CFD partial wave.  
Once more, even though our $\chi^2/d.o.f.$ does not have a well-defined statistical meaning, it  ensures  a nice description of the input as seen in the $\hat\chi^2\equiv\chi^2/d.o.f.$ values, given in Table~\ref{tab:chi11}. They come out $\hat\chi^2\simeq 1$ or less in all regions (we follow the same notation as for the S0 wave).

\begin{table}[!hbt] 
\caption{ \small Results in terms of $\hat\chi^2$ of the P-wave solutions I, II and III, in different regions}
\centering 
\begin{tabular}{c | c  c  c | c c} 
\hline\hline
 & $\hat\chi^2_{1}$ & $\hat\chi^2_{2}$ & $\hat\chi^2_{\IC}$ & $\hat\chi^2_{\delta}$ & $\hat\chi^2_{\eta}$\\
\hline\hline  
\rule[-0.05cm]{0cm}{.35cm} Solution I  & 0.7 & 0.3 & 0.3 & 1.3 & 0.9\\
\rule[-0.05cm]{0cm}{.35cm}  Solution II  & 0.7 & 0.2 & 0.4 & 1.2 & 1.3\\
\rule[-0.05cm]{0cm}{.35cm}  Solution III  & 0.7 & 0.4 & 0.4 & 1.6 & 0.4\\
\hline\hline
\end{tabular} 
\label{tab:chi11} 
\end{table}

\begin{figure}
\centering
\includegraphics[width=0.45\textwidth]{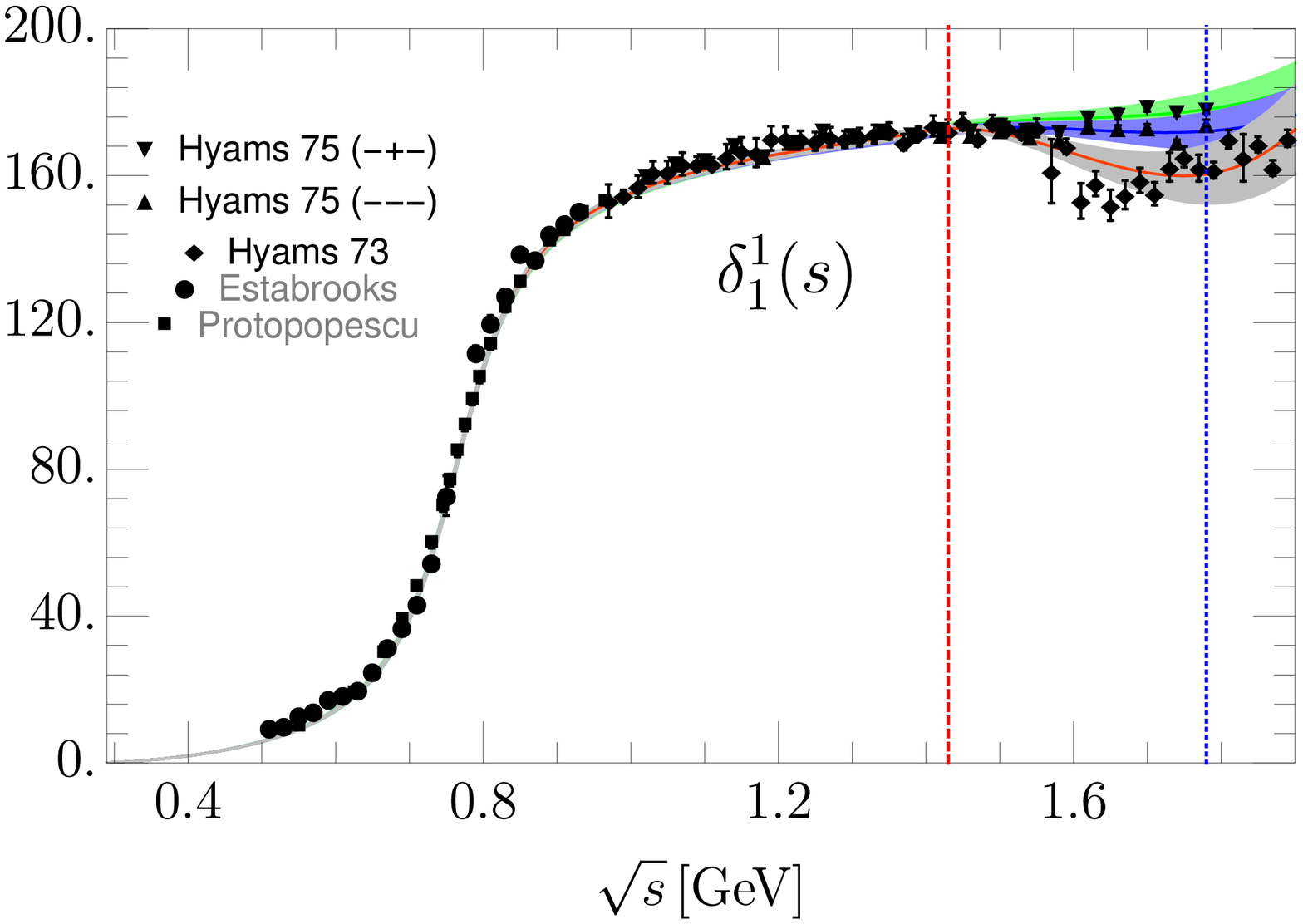}\vspace{.3cm} \\
\includegraphics[width=0.45\textwidth]{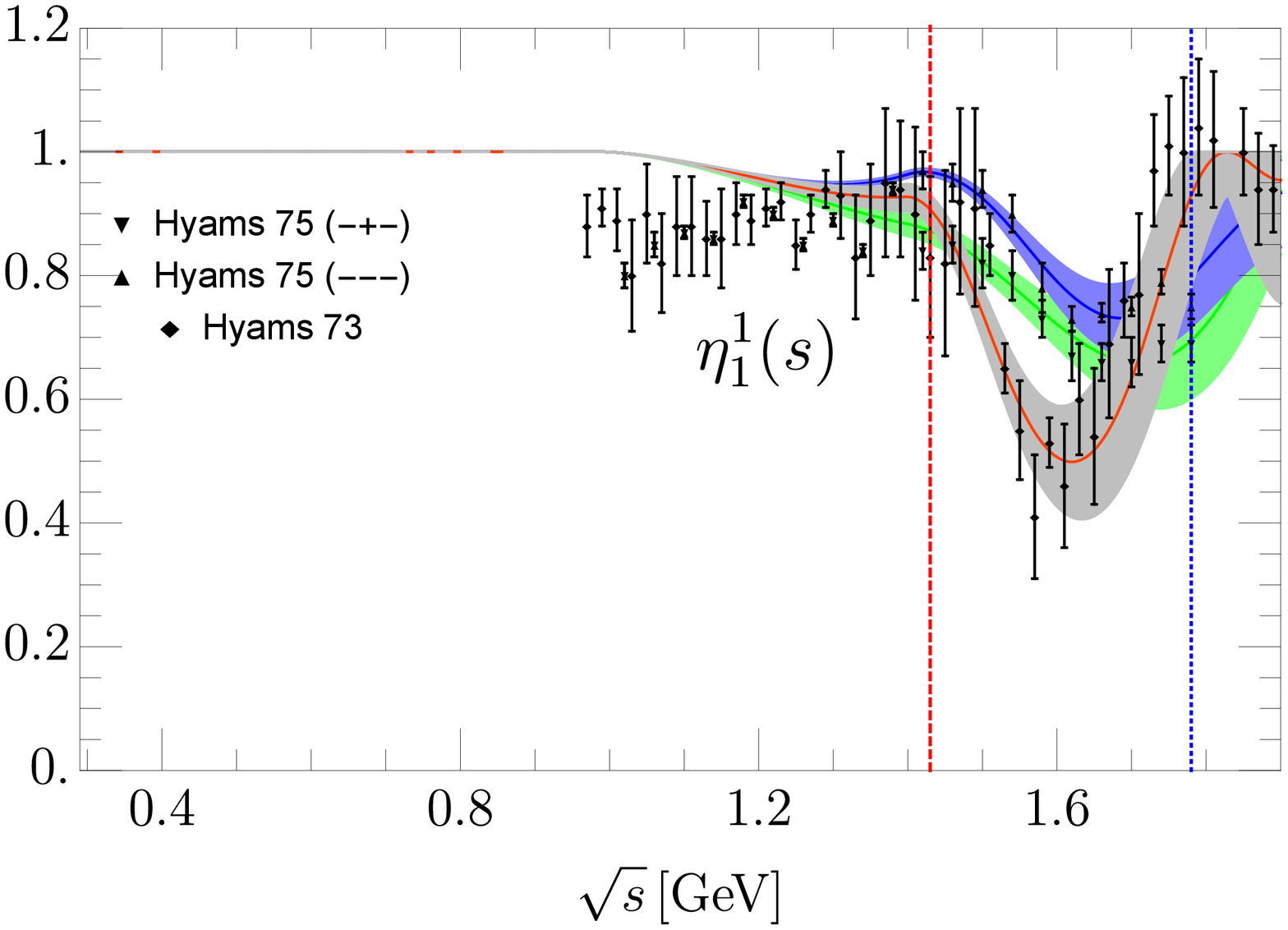}
\caption{\small\label{fig:shift-inelas11} Comparison between our three P-wave solutions and scattering data. The gray, blue and green bands correspond to solution I, II and III, respectively. The red dashed vertical line separates the region where the fits describe both data and dispersive results, from the region above where the parameterization is just fitted to the data. Namely, above 1.4 GeV solution I is fitted to ~\cite{Hyams:1973zf,Grayer:1974cr} (solid diamonds), solution II to~\cite{Hyams:1975mc} (solid upward triangles) and solution III to \cite{Ochs:2013gi} (solid downward triangles). 
The data from ~\cite{Protopopescu:1973sh} (solid squares), ~\cite{Estabrooks:1974vu} (solid circles) are just shown for completeness.
The blue dotted vertical line depicts the energy of the last data point of solutions II and III.}
\end{figure}

The resulting P-wave phase shift and elasticity are plotted in Fig.~\ref{fig:shift-inelas11} and their parameters are collected in Tables~\ref{tab:para11}, \ref{tab:para112}  and \ref{tab:para113} for solutions I, II and III, respectively. 
Both from the figure and the tables we can see that they are almost identical up to 1.4 GeV.
The uncertainties are described by the gray band for solution I, the blue band for solution II, and the green band for solution III. Below $s_m$ the three of them reproduce nicely the uncertainties given in~\cite{GarciaMartin:2011cn}.

\begin{table}[!hbt] 
\caption{ \small Fit parameters of the Global parameterization for the $P$-wave solution I.}
\centering 
\begin{tabular}{c c | c c | c c} 
\hline\hline
\multicolumn{2}{c}{$t^1_{1,\text{conf}}$} & \multicolumn{2}{c}{$\eta^1_1$} & \multicolumn{2}{c}{$\sqrt{s}>1.4 \gev$} \\
\hline\hline  
\rule[-0.05cm]{0cm}{.35cm}$B_0$ & 0.96$\pm$0.01 & $K_0$ &   0.049$\pm$0.002 & $d_0$ &   12.7$\pm$1.7\\
\rule[-0.05cm]{0cm}{.35cm}$B_1$ & 0.09$\pm$0.03 & $K_1$ & -0.005$\pm$0.0003 & $d_1$ &   6.0$\pm$0.5   \\ 
\rule[-0.05cm]{0cm}{.35cm}$B_2$ & -0.07$\pm$0.08  & & &  &    \\ 
\rule[-0.05cm]{0cm}{.35cm}$B_3$ & 0.58$\pm$0.19 & &  & $\epsilon_2$ &   -0.129$\pm$0.033\\ 
\rule[-0.05cm]{0cm}{.35cm}$B_4$ &  1.39$\pm$0.38 & &  & $\epsilon_3$ &    0.323$\pm$0.013  \\ 
\rule[-0.05cm]{0cm}{.35cm}$m_\rho$ &  0.7749$\pm$0.0012 \gev& & & $\epsilon_4$ &  0.200$\pm$0.007 \\ 
\hline\hline
\end{tabular} 
\label{tab:para11} 
\end{table}

\begin{table}[!hbt] 
\caption{ \small Fit parameters of the Global parameterization  for the $P$-wave  solution II.}
\centering 
\begin{tabular}{c c | c c | c c} 
\hline\hline
\multicolumn{2}{c}{$t^1_{1,\text{conf}}$} & \multicolumn{2}{c}{$\eta^1_1$} & \multicolumn{2}{c}{$\sqrt{s}>1.4 \gev$} \\
\hline\hline  
\rule[-0.05cm]{0cm}{.35cm}$B_0$ & 0.96$\pm$0.01 & $K_0$ &   0.054$\pm$0.001 & $d_0$ &   3.4$\pm$1.1\\
\rule[-0.05cm]{0cm}{.35cm}$B_1$ & 0.09$\pm$0.03 & $K_1$ & -0.0060$\pm$0.0001 & $d_1$ &   2.1$\pm$0.3   \\ 
\rule[-0.05cm]{0cm}{.35cm}$B_2$ & -0.03$\pm$0.08  & & &  &    \\ 
\rule[-0.05cm]{0cm}{.35cm}$B_3$ & 0.64$\pm$0.19 & &  & $\epsilon_2$ &   -0.14$\pm$0.02\\ 
\rule[-0.05cm]{0cm}{.35cm}$B_4$ &  1.1$\pm$0.32 & &  & $\epsilon_3$ &    0.041$\pm$0.005  \\ 
\rule[-0.05cm]{0cm}{.35cm}$m_\rho$ &  0.7749$\pm$0.0012 \gev& & & $\epsilon_4$ &  0.081$\pm$0.002 \\ 
\hline\hline
\end{tabular} 
\label{tab:para112} 
\end{table}

\begin{table}[!hbt] 
\caption{ \small Fit parameters of the Global parameterization  for the $P$-wave  solution III.}
\centering 
\begin{tabular}{c c | c c | c c} 
\hline\hline
\multicolumn{2}{c}{$t^1_{1,\text{conf}}$} & \multicolumn{2}{c}{$\eta^1_1$} & \multicolumn{2}{c}{$\sqrt{s}>1.4 \gev$} \\
\hline\hline  
\rule[-0.05cm]{0cm}{.35cm}$B_0$ & 0.96$\pm$0.01 & $K_0$ &   0.045$\pm$0.002 & $d_0$ &   3.1$\pm$1.0\\
\rule[-0.05cm]{0cm}{.35cm}$B_1$ & 0.07$\pm$0.03 & $K_1$ & -0.0037$\pm$0.0002 & $d_1$ &   1.7$\pm$0.3   \\ 
\rule[-0.05cm]{0cm}{.35cm}$B_2$ & -0.03$\pm$0.08  & & &  &    \\ 
\rule[-0.05cm]{0cm}{.35cm}$B_3$ & 0.67$\pm$0.18 & &  & $\epsilon_2$ &   -0.33$\pm$0.02\\ 
\rule[-0.05cm]{0cm}{.35cm}$B_4$ &  1.0$\pm$0.34 & &  & $\epsilon_3$ &    -0.105$\pm$0.007  \\ 
\rule[-0.05cm]{0cm}{.35cm}$m_\rho$ &  0.7749$\pm$0.0012 \gev& & & $\epsilon_4$ & $\equiv$ 0  \\ 
\hline\hline
\end{tabular} 
\label{tab:para113} 
\end{table}

Concerning the region below 1.4 GeV, note that the CFD uncertainties
are extremely small below $K\bar K$ threshold. 
For this reason, in order to ensure an accurate description of the error band in this region, we chose  $\alpha=0.3$ in Eq.~\eqref{eq:generalconformalP} so
that the center of the conformal expansion in our new amplitude is close to the \pipi threshold. In this way, the uncertainties there are dominated by the lowest conformal parameters $B_0$ and $B_1$, ensuring that the values of the threshold parameters, given in Table~\ref{tab:lowpara11}, are also consistent with the dispersive values in~\cite{GarciaMartin:2011cn}.
In contrast, the new parameterization uncertainties close to 1.4 $\gev$ are smaller than those of the CFD in~\cite{GarciaMartin:2011cn}, which is a consequence of  describing simultaneously the experimental data up to 2 \gev.  The P-wave elasticity in~\cite{GarciaMartin:2011cn} is compatible with 1 below 1.12 $\gev$ and slightly smaller below 1.4 $\gev$. This is why it can be reproduced by using only two free constants in the parameterization given in Eq.~\eqref{eq:ineP}.

The $\rho(770)$-pole parameters are given in Table~\ref{tab:poles11} and are identical for all solutions. Central values and uncertainties are nicely compatible with the dispersive results in~\cite{GarciaMartin:2011jx}.

\begin{table}[!hbt] 
\caption{ \small P-wave threshold parameters in customary $m_\pi=1$ units. They are almost indistinguishable for solutions I, II and III.}
\centering 
\begin{tabular}{c | c | c} 
\hline\hline
 & This work & Dispersive result \cite{GarciaMartin:2011cn} \\
\hline\hline  
\rule[-0.05cm]{0cm}{.35cm} $a^1_1$(x10$^3$) & 38.3$\pm$0.6  & 38.1$\pm$0.9 \\
\rule[-0.05cm]{0cm}{.35cm} $b^1_1$(x10$^3$) & 4.54$\pm$0.51  & 5.37$\pm$0.14 \\
\hline\hline
\end{tabular} 
\label{tab:lowpara11} 
\end{table}

Furthermore, as we already commented for the S-wave, this P-wave global parameterization up to 1.4 GeV is fully consistent with all the dispersion relations described in the the GKPY dispersive analysis ~\cite{GarciaMartin:2011cn}.
Since such a calculation requires as input the other partial waves and high-energy input described in ~\cite{GarciaMartin:2011cn}, we have relegated it to appendix \ref{app:dr}.

Above the matching point at 1.4 \gev the three solutions coming from the CERN-Munich experiment are incompatible among themselves. The behavior of solution I suggests a strong interference between the $\rho'$ and $\rho''$, with a sizable phase change around 1.6 \gev and a dip structure in the elasticity at the same energy. In contrast, solutions II and III look smoother. Namely, the phase grows slowly above $180^{\degree}$ and the elasticity has a less pronounced dip. In addition, the uncertainties quoted in~\cite{Hyams:1975mc} and \cite{Ochs:2013gi} are slightly smaller, which leads to more constrained uncertainty bands. Nevertheless we emphasize once more that above 1.4 GeV, we consider our parameterizations purely phenomenological.

\begin{table}[!hbt] 
\caption{ \small Pole position and $\pi\pi$ coupling of the $\rho(770)$, which are almost indistinguishable for solutions I, II and III. They nicely agree with the
GKPY dispersive result in~\cite{GarciaMartin:2011jx} that we also provide for comparison.}
\centering 
\begin{tabular}{c | c | c} 
\hline\hline
 & $\sqrt{s_{pole}}$ (MeV) & $|g|$ \\
\hline\hline  
\rule[-0.05cm]{0cm}{.35cm} $\rho(770)^{\text{GKPY}}$  & $(763.7^{+1.7}_{-1.5})-i (73.2^{+1.0}_{-1.1})$ & $6.01^{+0.04}_{-0.07}$\\
\rule[-0.05cm]{0cm}{.35cm} $\rho(770)$ & $(763.1\pm 1.5)-i (73.3\pm 1.4)$ & $5.99\pm0.06$\\
\hline\hline
\end{tabular} 
\label{tab:poles11} 
\end{table}



\section{Summary}

In this work we have provided a global parameterization of the data for each one of the S0 and P-waves of \pipi scattering up to almost 2 GeV. We have made an explicit effort to keep it relatively simple in order to be easy to implement in further phenomenological and experimental analyses (in final state interactions, isobar models, etc...).

The advantages of these parameterizations are that they describe experimental data up to 2 GeV consistently with the dispersive representation in~\cite{GarciaMartin:2011cn}
and its uncertainties up to its maximum applicability region of 1.4 GeV. In addition, they reproduce the dispersive results within their applicability region in the complex-$s$ plane, as obtained in~\cite{GarciaMartin:2011jx}, including the poles associated to the \sig, $f_0(980)$ and $\rho(770)$ resonances. Moreover, their low-energy behavior is compatible with the dispersive results for the threshold parameters and the S0 Adler zero and hence with the constraints due to the QCD spontaneous chiral symmetry breaking.

Actually, these new parameterizations reproduce the results and uncertainties of a previous piecewise fit that was constrained to satisfy Forward Dispersion relations up to 1.4 GeV and partial-wave dispersion relations (Roy and GKPY equations) up to 1.12 GeV. The latter were used in~\cite{GarciaMartin:2011cn} to obtain a rigorous analytic continuation to the complex plane which, together with its uncertainties, is also described when continuing analytically our new parameterization, without the need for a numerical integration of the dispersion relations. This is why the pole positions and residue of the \sig, the $f_0(980)$ and the $\rho(770)$ are so well implemented. It also allows our parameterization to be used consistently in applications with isobar models, so popular in experimental analyses.

The new parameterizations also reproduce the existing data from 1.43 to 2 GeV, although the dispersion relations do not reach these energies. Moreover, in this region, there are three contradictory data sets, and we thus provide three solutions for each wave that describe phenomenologically either one of the conflicting sets. Nevertheless, below 1.4 GeV these three solutions agree to the point of being almost indistinguishable and are consistent with the dispersive analysis.

We hope that the simplicity and the remarkable analytic properties of this data parameterization can be useful for future phenomenological and experimental studies in which \pipi interactions are needed.

\begin{acknowledgments} 
 We thank prof. W. Ochs for his comments and for providing us with the data of his updated $(- + -)$ solution. JRP and AR are supported by the Spanish project FPA2016-75654-C2-2-P. This project has received funding from the European Union’s Horizon 2020 research and innovation program under grant agreement No.824093. AR would also like to acknowledge the financial support of the Universidad Complutense de Madrid through a predoctoral scholarship.  JRE  is supported by the Swiss National Science Foundation, project No.\ PZ00P2\_174228.
\end{acknowledgments}

\appendix
\section{Roy-Steiner and Forward dispersion relations}
\label{app:dr}

As we have explained in the main text, the global parameterizations we present in this work 
mimic very well the dispersive piece-wise CFD parameterization of \cite{GarciaMartin:2011cn}  and their uncertainties up to almost 1.4 GeV. Of course, they are not exactly identical to the CFD and one might wonder if the dispersion relations used in \cite{GarciaMartin:2011cn} to constrain the piece-wise parameterization are still satisfied. In this appendix we show that this is indeed the case.

Here we only show the results for solution I since, as we have emphasized repeatedly, up to 1.4 GeV all our three solutions are almost indistinguishable.
The explicit equations for each dispersion relation as well as the rest of the input apart from the S0 or P-waves below 1.4 GeV can be found in \cite{GarciaMartin:2011cn} and references therein. For the S0 and P waves below 1.4 GeV we use the expressions in the main text here. Since our DR only reach up to 1.4 GeV at most, the results with the other solutions are indistinguishable.

Let us simply recall that each dispersion relation provides the real part of the amplitude as an integral from threshold to infinity of the imaginary part times some specific kernel for each relation. Sometimes some subtraction terms (polynomials in $s$), which are real, are  added to this integral, and their coefficients (subtraction constants) are obtained from the values of the amplitude at fixed values, typically at or around threshold. The imaginary part is taken from the fits to data. That integral and the subtraction terms provide what we call the ``Dispersive'' result. When it agrees within uncertainties with the values of the amplitude obtained from our fits, we consider that the dispersion relation is well satisfied.

\begin{figure}
\begin{center}
\includegraphics[width=0.47\textwidth]{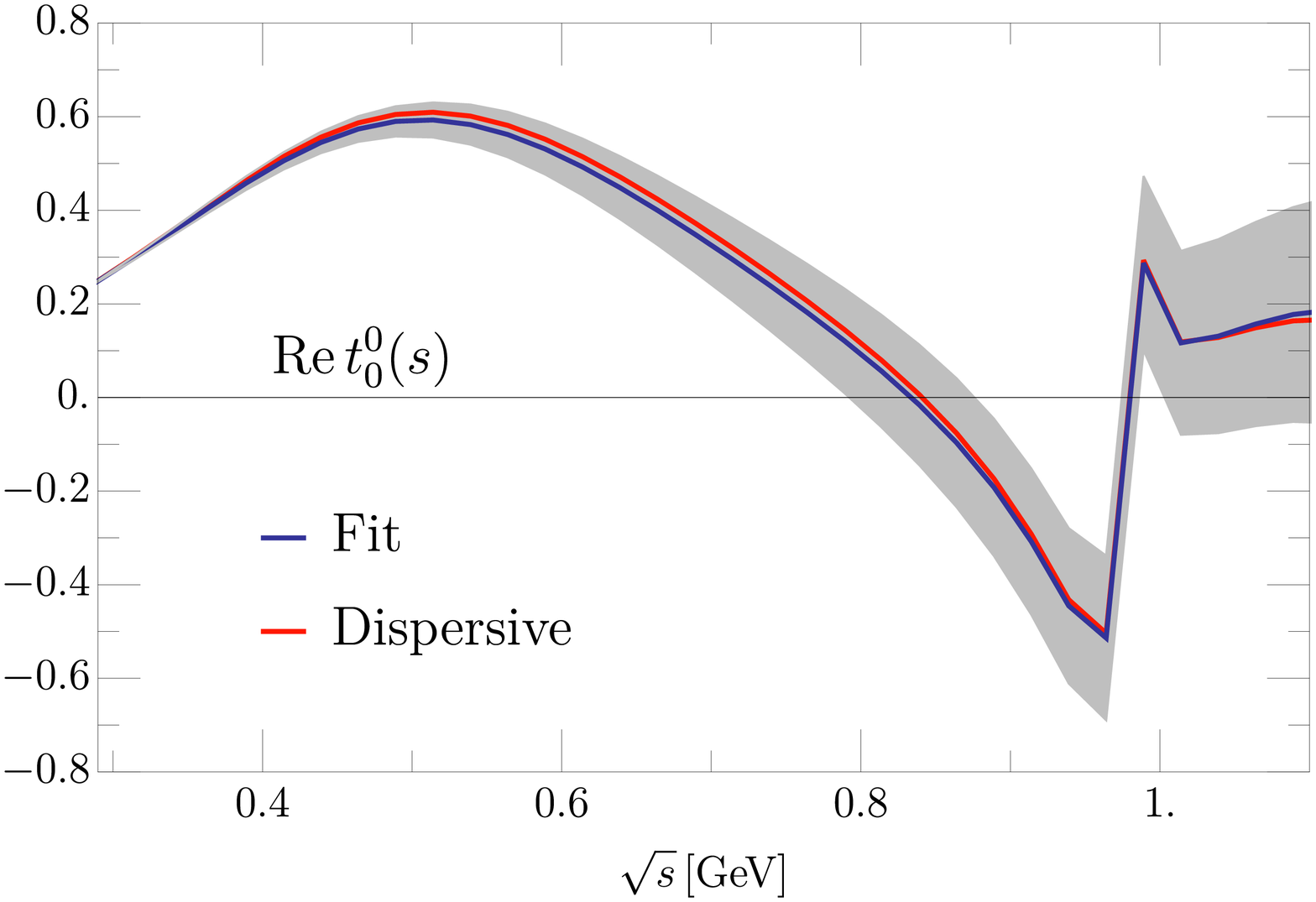}

\vspace{.2cm}
\includegraphics[width=0.47\textwidth]{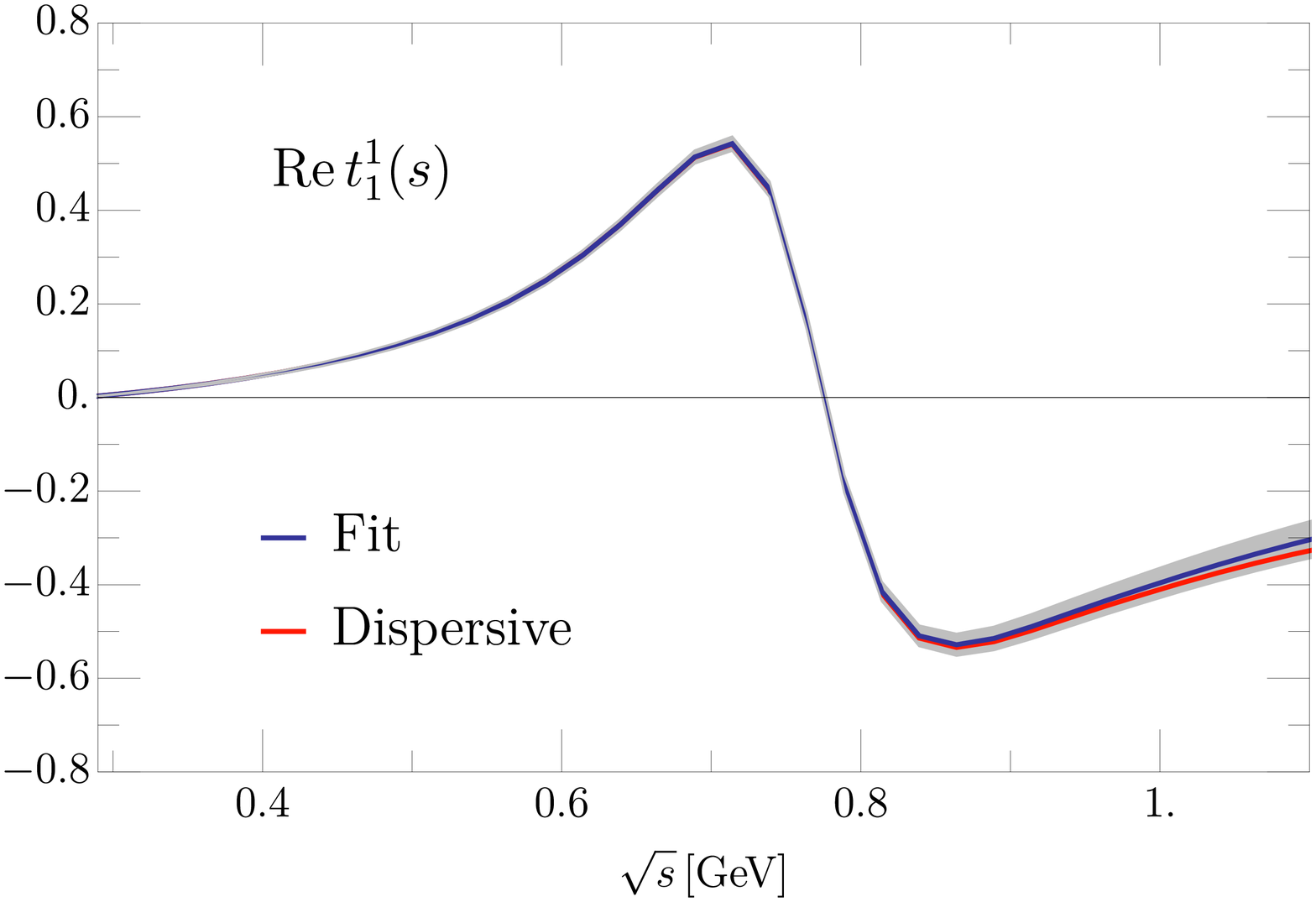}

\vspace{.2cm}
\includegraphics[width=0.47\textwidth]{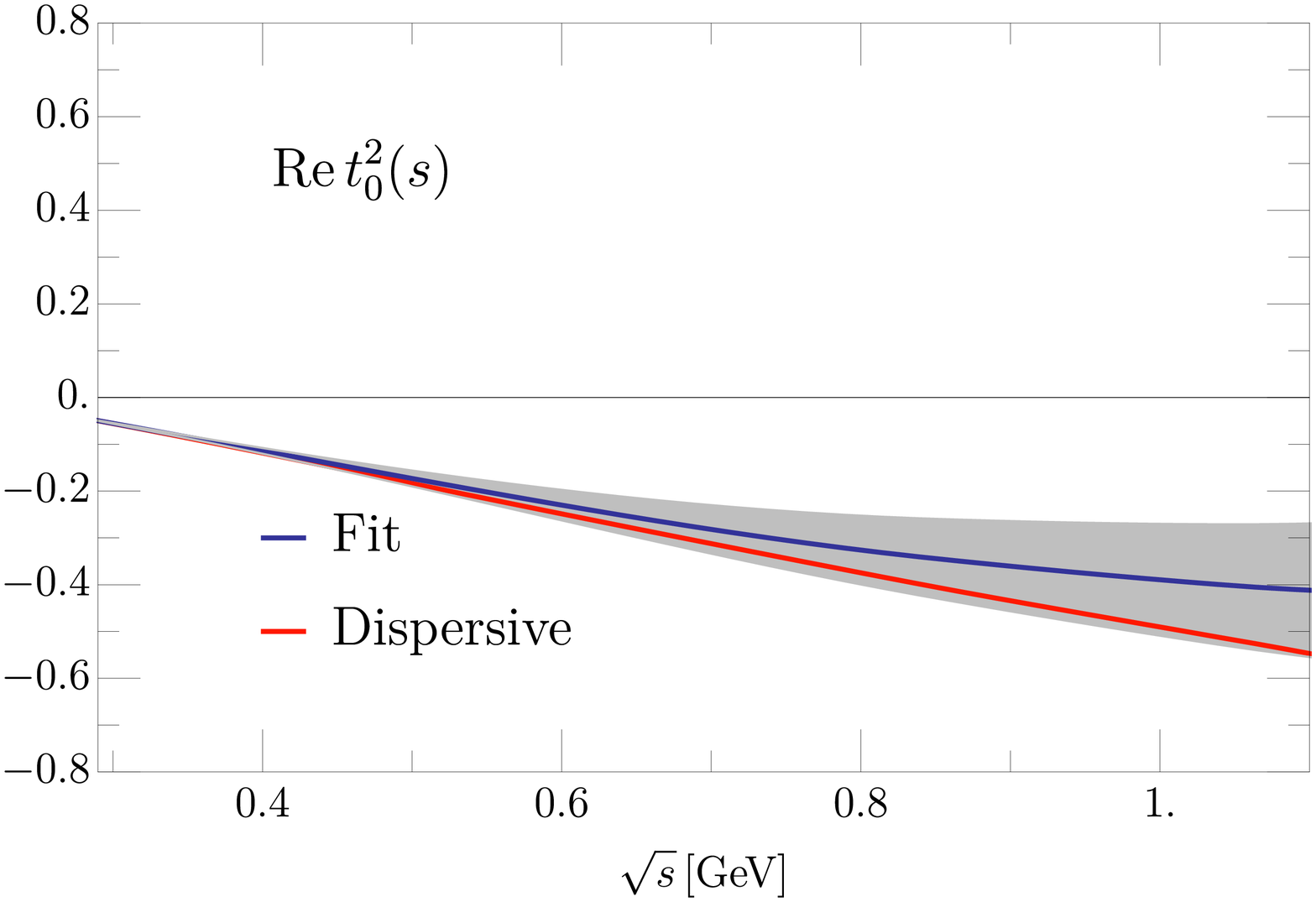}
\end{center}
\vspace*{-18pt}
\caption{ \small Results for Roy equations. Blue lines: real part coming from our new parameterizations. Orange lines: the result of the dispersive integrals. The gray bands cover the uncertainties in the difference between both. From top to bottom: (a) $S0$ wave, (b) $P$ wave, (c) $S2$ wave.
\label{fig:CFD-Roy-plots}}
\vspace*{-10pt}
\end{figure}

\begin{figure}
\begin{center}
\includegraphics[width=0.47\textwidth]{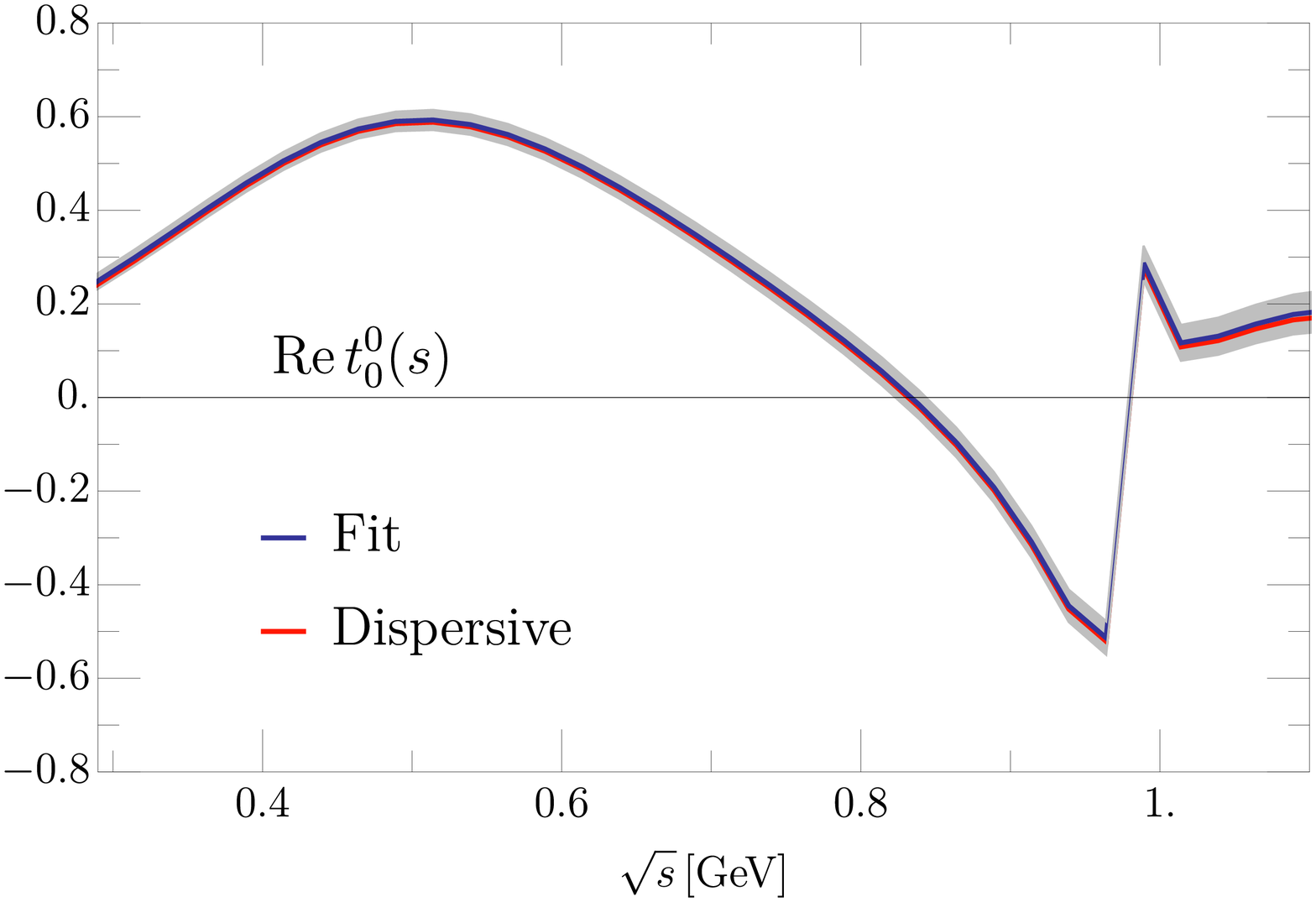}

\vspace{.2cm}
\includegraphics[width=0.47\textwidth]{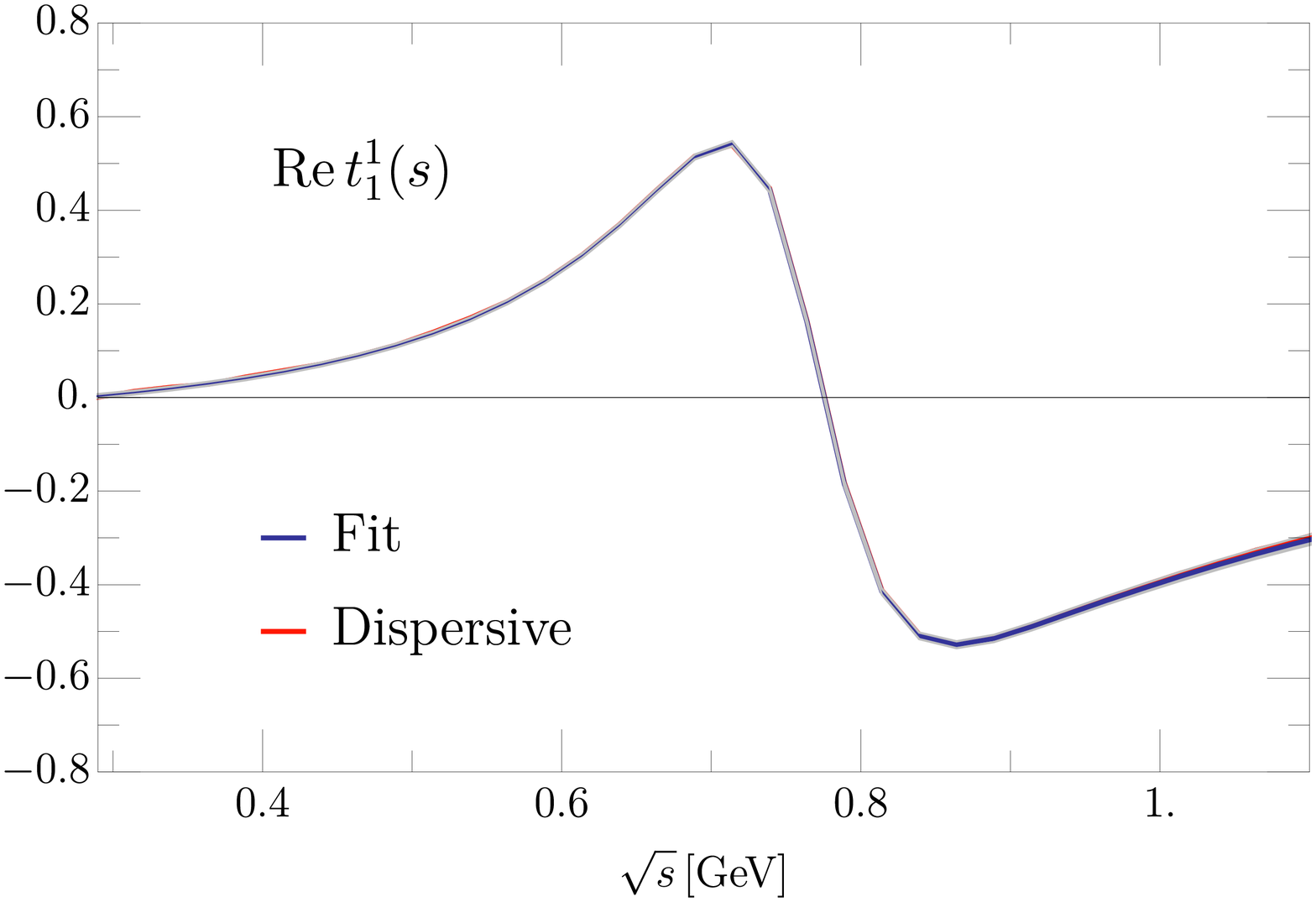}

\vspace{.2cm}
\includegraphics[width=0.47\textwidth]{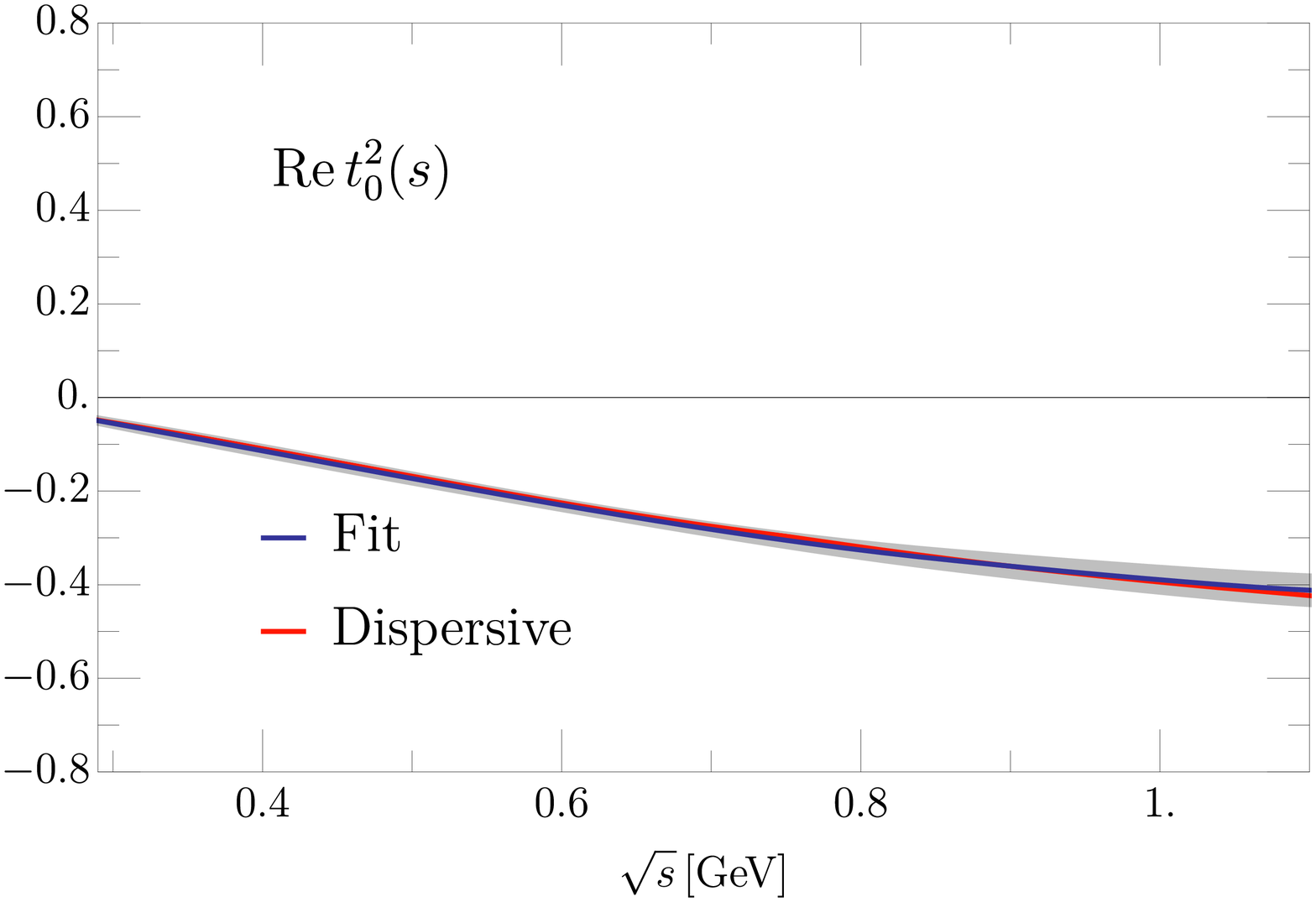}
\end{center}
\vspace*{-18pt}
\caption{ \small Results for GKPY equations. Blue lines: real part coming from our new parameterizations. Orange lines: the result of the dispersive integrals. The gray bands cover the uncertainties in the difference between both. From top to bottom: (a) $S0$ wave, (b) $P$ wave, (c) $S2$ wave.
\label{fig:CFD-GKPY-plots}}
\vspace*{-10pt}
\end{figure}

Thus, in Fig.~\ref{fig:CFD-Roy-plots} we show the ``Dispersive'' result versus our fit for the the S0, P and S2 Roy equations. It can be seen that in all cases they are well satisfied within uncertainties within the whole applicability region up to $\sqrt{s}\sim 1.1\,$GeV. The same can be said about the GKPY equations, which are similar to Roy equations but with just one subtraction. They are more sensitive to the resonance region and less to the threshold region. In any case, it can be observed in Fig.~\ref{fig:CFD-GKPY-plots} that they are also well satisfied within uncertainties within their applicability region.

\begin{figure}
\begin{center}
\includegraphics[width=0.47\textwidth]{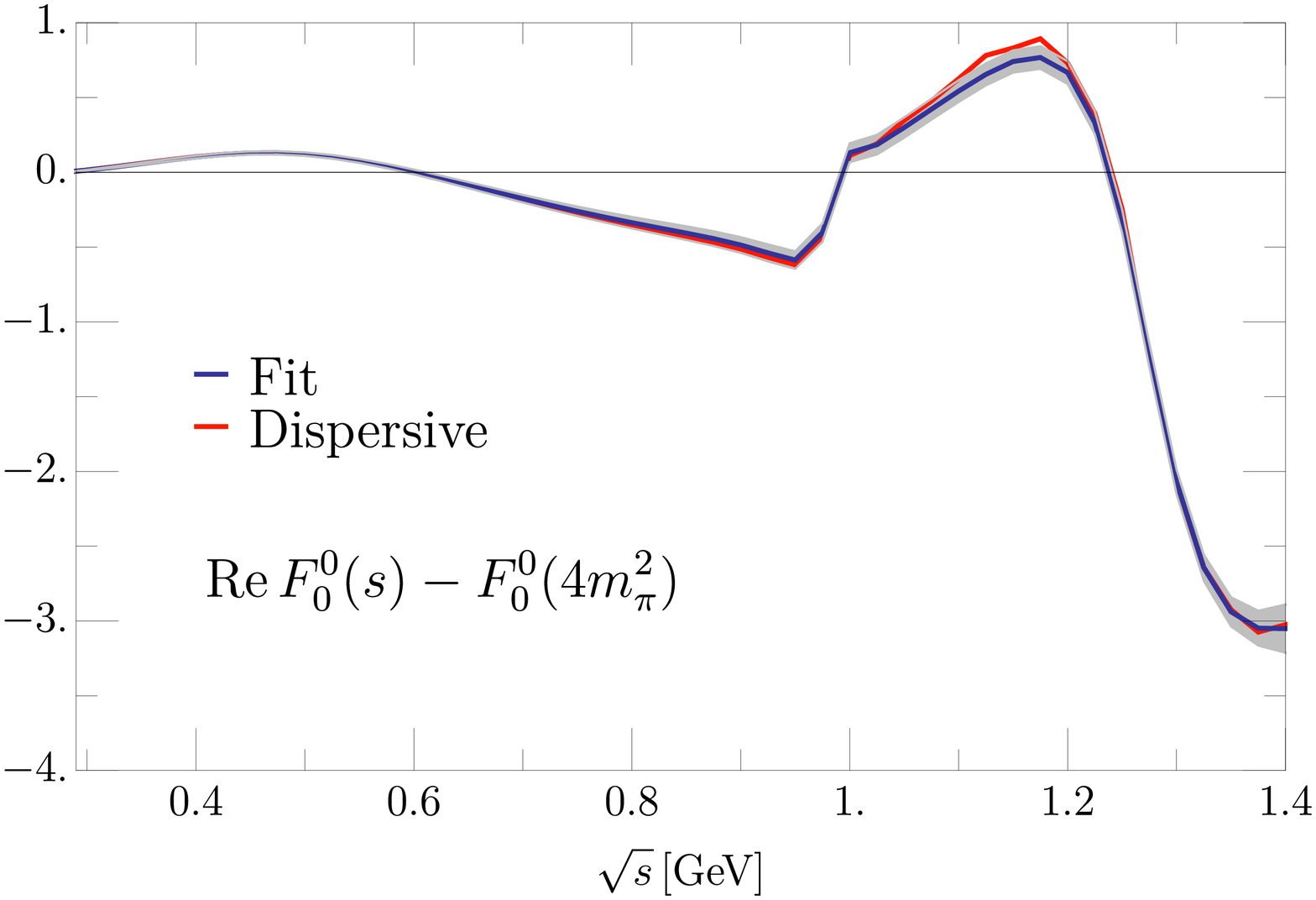}

\vspace{.2cm}
\includegraphics[width=0.47\textwidth]{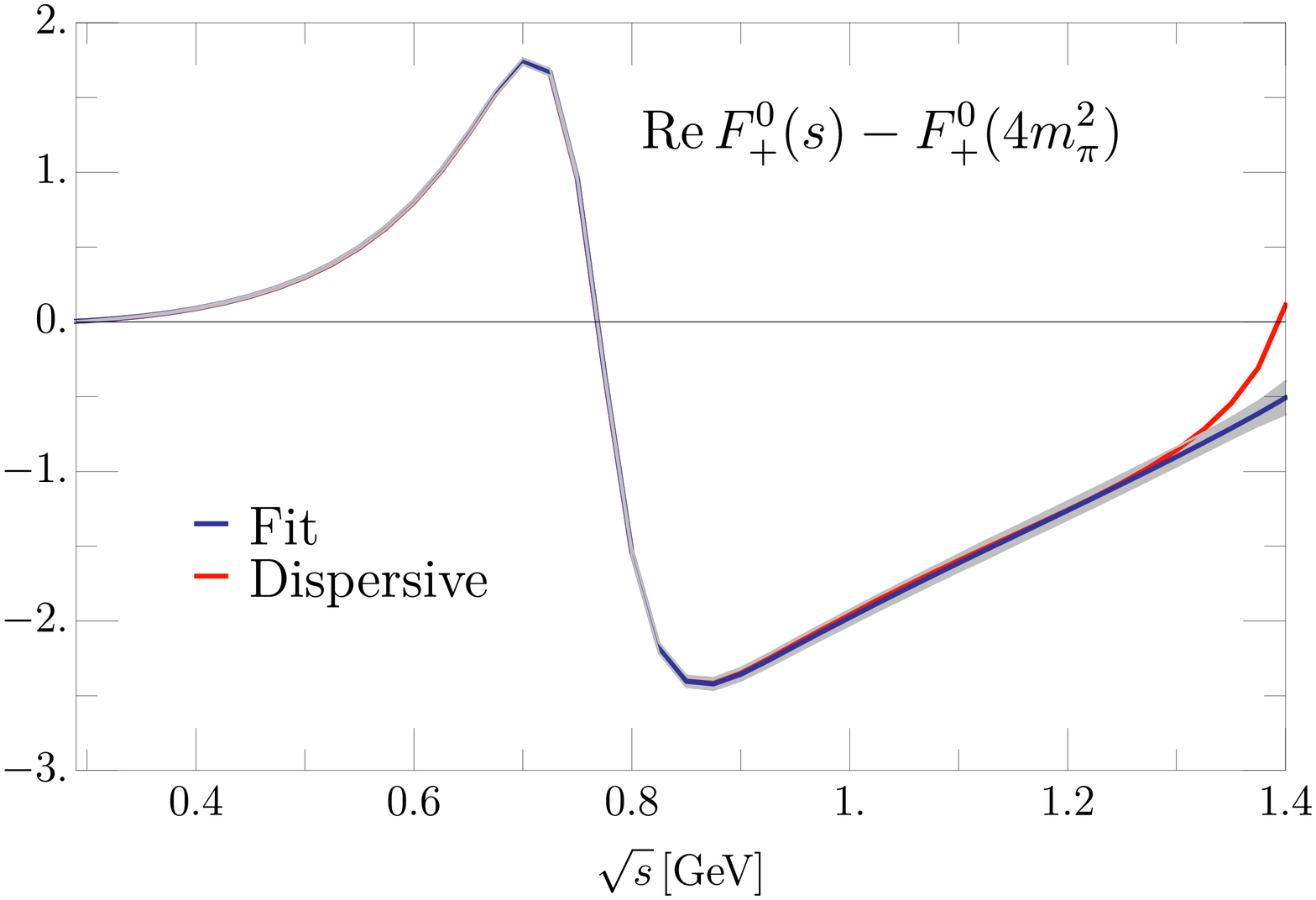}

\vspace{.2cm}
\includegraphics[width=0.47\textwidth]{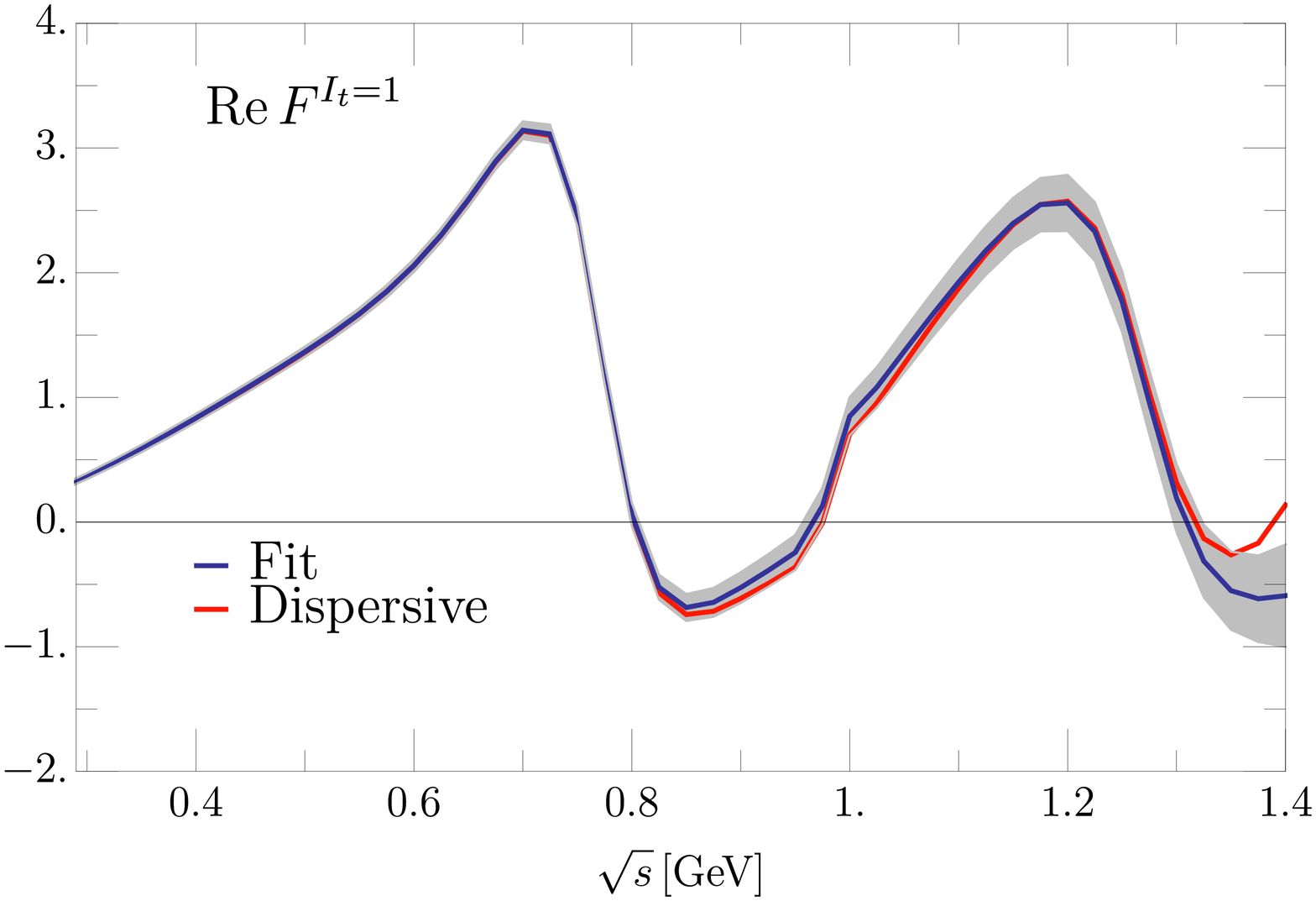}
\end{center}
\vspace*{-18pt}
\caption{ \small Results for forward dispersion relations. Blue lines: real part coming from our new parameterizations. Orange lines: the result of the dispersive integrals. The gray bands cover the uncertainties in the difference between both. From top to bottom: (a) the $\pi^0\pi^0$ FDR, (b) the $\pi^0\pi^+$ FDR, (c) the FDR for $I_t=1$ scattering.
\label{fig:CFD-FDR-plots}}
\vspace*{-10pt}
\end{figure}

Finally, we show in Fig.~\ref{fig:CFD-FDR-plots} the fulfillment of Forward Dispersion Relations (FDR). It should be noted that these are not written for partial waves, but for the amplitudes
\begin{eqnarray}
  \label{fdr:basis}
  F^{0}_0&=&\frac{1}{3}(F^{(0)}+2F^{(2)}),\quad
  F^{0}_+=\frac{1}{2}(F^{(1)}+F^{(2)}),\nonumber \\
  F^{I_t=1}&=&\frac{1}{3}F^{(0)}+\frac{1}{2}F^{(1)}-\frac{5}{6}F^{(2)}.\nonumber 
\end{eqnarray}
where
\begin{equation}
F^{(I)}(s,t)=\frac{8}{ \pi} \sum_\ell (2\ell+1)P_\ell(\cos\theta) t^{I}_\ell(s).
\end{equation}
Once again we see that these FDR are well satisfied up to 1.4 GeV, which is the energy up to which they were studied before in \cite{GarciaMartin:2011cn}. Above that energy, Regge theory is being used as input for the integrals instead of partial waves. Note that  in \cite{GarciaMartin:2011cn} the FDRs were checked up to 1.42 GeV whereas here we only check them up to 1.4 GeV since it is at that point that we make the matching with the different solutions of higher energies and our new global parameterization starts deviating from our piece-wise solution to accommodate the different data sets above 1.4 GeV. 
As seen in Fig.~\ref{fig:CFD-FDR-plots} the agreement is remarkable except in the final points very close to 1.4 GeV, i.e, the matching point, where our global fit has to deviate slightly from the old CFD to match the data above 1.4 GeV. Nevertheless, note that, even in  \cite{GarciaMartin:2011cn} there was a small deviation in the FDR around 1.4 GeV due to the discontinuous matching with the Regge description. 

Moreover, following \cite{GarciaMartin:2011cn} and to quantify the agreement between the fit and the dispersive result we have defined $\Delta_i(s)$ as the difference
between the dispersive output and the fit for each dispersion relation $i$,
whose uncertainties are called $\delta\Delta_i(s)$. 
Next, the average discrepancies are defined as follows:
\begin{equation}
\bar{d}_i^2\equiv\frac{1}{\hbox{number of points}}
\sum_n\left(\frac{\Delta_i(s_n)}{\delta\Delta_i(s_n)}\right)^2 ,
\label{avdiscrep}
\end{equation}
where the values of $\sqrt{s_n}$ are taken at intervals of 25 MeV as in~\cite{GarciaMartin:2011cn}. These discrepancies are similar in form to a $\chi^2$ function
and when they are $\sim1$ or smaller it means that the dispersion relation is well satisfied.
The resulting discrepancies are shown in Table~\ref{tab:CFDdiscrepancies} for different energy intervals. The fulfillment of these dispersion relations is remarkable, since all the average discrepancies are less than one except for the 1.25 value of the $\pi^+\pi^0$ up to 1.4 GeV. Even that is very good taking into account that the uncertainties of the global fit  are somewhat smaller close to 1.4 GeV than for the old CFD piece-wise parameterization, due to the matching with additional data above that energy.
Actually these discrepancies are on the average very similar in size to those obtained in~\cite{GarciaMartin:2011cn} with the piece-wise CFD parameterizations. 
The consistency of our global parameterization with respect to dispersion relations is thus remarkable and comparable to the one in~\cite{GarciaMartin:2011cn}.

\begin{table}
\begin{tabular}{|l|c|c|}
\hline
FDRs&{\small $s^{1/2}\leq 932\,$MeV}&{\small $s^{1/2}\leq 1400\,$MeV}\\
\hline
$\pi^0\pi^0$& 0.14 & 0.36\\
$\pi^+\pi^0$& 0.10 & 1.25\\
$I_{t=1}$& 0.12 & 0.29\\
\hline
\hline
Roy Eqs.& $s^{1/2}\leq 992\,$MeV & $s^{1/2}\leq 1100\,$MeV \\
\hline
S0& 0.13&0.11\\
S2& 0.44&0.50\\
P& 0.03& 0.07\\
\hline
\hline
GKPY Eqs.& $s^{1/2}\leq 992\,$MeV & $s^{1/2}\leq 1100\,$MeV \\
\hline
S0& 0.06&0.06\\
S2& 0.19&0.17\\
P& 0.21& 0.25\\
\hline
\hline
Average&0.16&0.34\\
\hline 
\end{tabular}
\caption{ \small  Average discrepancies $\bar d_i^2$ of our global parameterizations for each dispersion relation. \label{tab:CFDdiscrepancies}}
\end{table}

As a final comment, note that if we had neglected the P-wave tiny inelasticity between 1 GeV and 1.12 GeV, setting $\eta^1_1=1$ there, then $\bar d_{\pi^+\pi^0}^2$ up to 1.4 GeV would have been $\sim 6$. It is for this reason that we have have chosen to introduce $t^1_{1,in}$ in our P-wave parameterization, Eq.~\eqref{eq:finalP}.

\bibliographystyle{h-physrev}
\bibliography{largebiblio.bib}
\end{document}